\documentclass[useAMS,usenatbib]{mn2e}
\usepackage{epsfig}
\usepackage{natbib}
\usepackage{soul}
\title[The Slope of the Near Infrared Extinction Law]{The Slope of the Near Infrared Extinction Law}
\author[]{J.J. Stead$^{1}$\thanks{E-mail:
phy2j2s@leeds.ac.uk; mgh@leeds.ac.uk}, M.G. Hoare$^{1}$\footnotemark[1]\\
$^{1}$School of Physics and Astronomy, University of Leeds, Leeds, LS2 9JT\\
}

\def\Bes{\rm Besan$\c{c}$\rm on}
\def\arcmin{^{\prime}}
\begin{document}

\date{}

\pagerange{\pageref{firstpage}--\pageref{lastpage}} \pubyear{2002}

\maketitle

\label{firstpage}

\begin{abstract}
We determine the slope of the near infrared extinction power law (A$_{\lambda} \propto \lambda^{-\alpha}$) for 8 regions of the Galaxy between l$\sim$27$^{\circ}$ and $\sim$100$^{\circ}$.  UKIDSS Galactic Plane Survey data are compared, in colour-colour space, with Galactic population synthesis model data reddened using a series of power laws and convolved through the UKIDSS filter profiles. Monte Carlo simulations allow us to determine the best fit value of $\alpha$ and evaluate the uncertainty. All values are consistent with each other giving an average extinction power law of $\alpha$=2.14$^{+0.04}_{-0.05}$. This is much steeper than most laws previously derived in the literature from colour excess ratios, which are typically between 1.6 and 1.8. We show that this discrepancy is due to an inappropriate choice of filter wavelength in conversion from colour excess ratios to $\alpha$ and that effective rather than isophotal wavelengths are more appropriate. In addition, curved reddening tracks, which depend on spectral type and filter system, should be used instead of straight vectors.
\end{abstract}


\section{Introduction}
\label{sec:intro}

Due to its affect on most astronomical observations, interstellar extinction remains an important issue for the study of virtually all objects. Stars positioned towards the Galactic Plane, particularly those inside or behind star forming regions, are affected most due to the increasing amounts of interstellar dust along the line of sight. The extinction suffered at visible wavelengths is typically an order of magnitude larger than that suffered in the K band. Furthermore, the wavelength dependence of interstellar extinction varies far more in the optical and ultraviolet than in the infrared, where it is commonly treated as universal and can be represented as power law (A$_{\lambda} \propto \lambda^{-\alpha}$) \citep{mathis90,draine03}. There are however noted cases in the literature of significant variance with flatter extinction laws arising, due to grain growth, in dense clouds \citep{he95,racca02}.\\  
$\indent$Due to the lessened effects of extinction, recent Galactic surveys now use infrared light to probe the densest parts of the Galaxy. Although the amount of extinction suffered is reduced, extinction correction is still required when attempting to deredden obscured sources. This is particularly true in star forming regions where there can be strong differential reddening between neighbouring cluster members. \\
$\indent$Both \citet{mathis90} and \citet{draine03} quote several values of $\alpha$, fitted by several authors, between 1.6 and 1.8. These values have mostly been derived by first measuring the near-infrared (NIR) colour excess ratio of well studied stars with known spectral types. NIR colour excess ratios can be converted to a value of $\alpha$ using the following relationship:
\begin{equation}
  \frac{E_{J-H}}{E_{H-K}} = \frac{(\frac{\lambda _{H}}{\lambda _{J}})^{\alpha}-1}{1-(\frac{\lambda _{H}}{\lambda _{K}})^{\alpha}},
  \label{eq:no1}
\end{equation}
where $\lambda _{J}$ is the assumed wavelength of the J photometric filter. \\
$\indent$It is commonly noted in the literature that the effective filter wavelengths are colour dependent. Using the above equation, we would derive different values of $\alpha$ for effective wavelengths computed from different colour spectra. However the difference between effective filter wavelengths derived from sources of differing spectral type and reddening has usually been considered negligible. With the advancement of deeper NIR surveys it is now possible to observe objects with highly reddened spectra and the previous assumption must be questioned. \\
$\indent$For highly reddened lines of sight in the Galactic Plane, individual stars with well known spectral types are rare. Despite this it is still possible to analyse the interstellar extinction using the red giant locus on colour-colour diagrams (CCDs). \citet{indebetouw05} isolated red clump giants to determine colour excess ratios and the wavelength dependence of interstellar extinction (A$_{\lambda}$/A$_K$) towards two different lines of sight. The first is noted as an unremarkable region, the second contains the giant HII region RCW 49, and a field unrelated to the HII region. They conclude that the wavelength dependence of interstellar extinction is remarkably similar despite the difference in Galactic location. \citet{marshall06} created a three dimensional extinction distribution model of the inner Galaxy using a catalogue of synthetic stars, created using the $\Bes$ stellar population synthesis model of the Galaxy \citep{robin03} assuming the extinction power law of \citet{mathis90}. \\
$\indent$In this paper we derive a NIR extinction power law along 8 different lines of sight, distributed along the Galactic Plane ($\mid$b$\mid$$<$5deg), each centred on an object in the Red MSX Source (RMS) survey \citep{urquhart07}. The RMS survey is an ongoing multi-wavelength observational programme designed to return a large, well-selected sample of massive young stellar objects (MYSOs) with the primary basis from the Midcourse Space Experiment (MSX). For this reason, towards the centre of each region of sky studied, there will be a massive star forming region, containing either an MYSO or an ultra-compact HII (UCHII) region. Therefore the data will also contain highly reddened objects, situated inside or behind the thick natal cloud of the massive star. However, since the regions of sky studied are large, the dominant cause of extinction will still be due to the diffuse ISM. \\
$\indent$Each source in the RMS survey has a computed kinematic distance, from $^{13}$CO observations \citep{urquhart08}, used to determine the luminosity of each MYSO candidate. This however is not without certain pitfalls. Kinematic distances are typically only accurate to $\sim$1kpc, and any RMS object positioned in the inner Galaxy will possess a near/far kinematic ambiguity. \citet{busfield06} attempted to solve this ambiguity, to varying degrees of success, using the HI self-absorption technique to determine the near or far distance. In order to calculate an independent distance to clusters associated with MYSOs, Stead \& Hoare (in prep.) fit isochrones to the intrinsic colours of individually dereddened cluster members. Although the amount of extinction suffered at infrared wavelengths is reduced, in comparison to optical wavelengths, extinction correction is still required when attempting to deredden obscured sources. This is particularly true in star forming regions where there can be strong differential reddening between neighbouring cluster members. Small changes in the extinction power law used to deredden individual objects can have a large effect the results. It is therefore necessary to determine an accurate value of $\alpha$ specific towards a particular line of sight. \\
$\indent$In section \ref{sec:data} we describe the NIR data used to derive the extinction power law. In section \ref{sec:redden} we discuss the implications of allowing the filter wavelengths of photometric systems to evolve with the progressively reddened observed spectra. In section \ref{sec:modelcdd} we describe the method used to derive a value of $\alpha$. In section \ref{sec:results} we derive NIR extinction power laws for the remaining 7 regions. We draw comparisons between our work and that of previous authors in section \ref{sec:discussion}. We summarise our conclusions in section \ref{sec:conclusion}.

\section{Data}
\label{sec:data}
We extract data for each region from two different NIR catalogues, the United Kingdom Infrared Deep Sky Survey Galactic Plane Survey (UKIDSS GPS) and 2MASS. 
\subsection{The UKIDSS GPS}
\begin{figure}
  \begin{center}
    \begin{tabular}{cc} \resizebox{80mm}{!}{\includegraphics[angle=0]{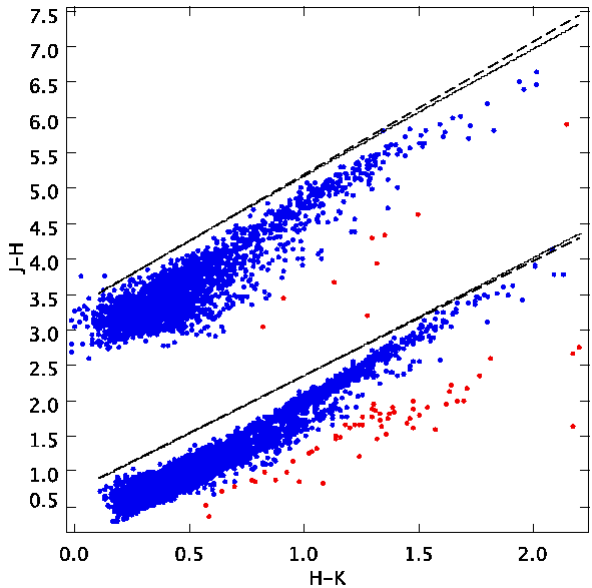}}
    \end{tabular}
    \caption{\small An example CCD of UKIDSS data (bottom) and 2MASS data (top). The 2MASS data have been offset along the J-H axis by 2.5. Due to the increased photometric depth of the UKIDSS data, the giant branch is more prominent compared with the 2MASS CCD. There are 6638 UKIDSS sources and 2694 2MASS sources extracted from the same region of sky, centred on the RMS source G48.9897-00.2992, covering an area of 60$^{\prime}$x6$^{\prime}$. The data cuts detailed in the text have been applied. The UKIDSS data (and 2MASS data to a lesser extent) contain many sources scattered below the bulk of the data (in each case highlighted in red). These sources most likely possess an IR excess due to emission from circumstellar dust. Such stars are abundant in star forming regions. Curved reddening tracks, created with a K0III spectrum and $\alpha$=2.1 specific to each photometric system, are plotted as solid black lines. Straight reddening vectors, given the same initial gradient as each corresponding reddening track, are overlaid as dashed lines.}
    \label{fig:exampleCCD}
  \end{center}
\end{figure}

The UKIDSS Galactic Plane Survey (GPS, see \citet{lucas08}) covers the region of the Galactic Plane accessible by the United Kingdom Infrared Telescope (UKIRT), in the J(1.248$\mu$m), H(1.631$\mu$m) and K(2.201$\mu$m) filters \citep{lawrence07}. The GPS data are obtained from the Wide Field Camera (WFCAM) Science Archive \citep{hambly08}. The median 5- $\sigma$ depths in the second data release of the GPS, given by \citet{warren07}, are J=19.77, H=19.00 and K=18.05 (Vega system). However the survey depth is spatially variable, typically decreasing longitudinally towards the Galactic centre and latitudinally towards the Galactic Plane as the fields become more crowded. \citet{lucas08} determine the modal depths in uncrowded fields to be J=19.4 to 19.65, H=18.5 to 18.75 and K=17.75 to 18.0. \\
$\indent$The WFCAM Science Archive enables several different types of data quality cuts to make it possible to select only the most reliable data. These include ellipticity cuts to remove blended objects, warnings to remove sources with saturated pixels, astrometry cuts to remove sources mismatched between different filters, and finally photometric error cuts. The data cuts used are the same as those presented in \citet{lucas08}. The only alteration we make is a photometric error cut of 0.03 mag in each photometric filter, whereas \citet{lucas08} apply a 0.03 mag colour cut. The use of such data quality cuts mean that no sources fainter than approximately J$\sim$18.0, H$\sim$16.8 and K$\sim$16.1 are present in the data. An example CCD created using UKIDSS data is presented in Fig.\ref{fig:exampleCCD}. The UKIDSS GPS data saturate at K$\la$12 and so we use 2MASS data to provide a way to analyse the brighter stars in each field.
\subsection{2MASS}
The second catalogue is the Two Micron All Sky Survey (2MASS, see \citet{skrutskie06}) covering the whole sky in the J(1.235$\mu$m), H(1.662$\mu$m) and K$_S$(2.159$\mu$m) filters \citep{cohen03}). The 2MASS data have median 3-$\sigma$ depths of J=17.1, H=16.4 and K$_S$=15.3. Only 2MASS sources with photometric errors in each band less than 0.1 mag have been selected. A larger photometric error cut was used on 2MASS data, in comparison to the UKIDSS data, as reducing the cut to that of the UKIDSS data (0.03 mag) drastically reduces the number of returned sources. The application of this photometric error cut means that no sources fainter than approximately J$\sim$16.5, H$\sim$15.3 and K$\sim$14.9 are present in the data. An example CCD created using 2MASS data is presented in Fig.\ref{fig:exampleCCD}.
\section{Effective filter wavelengths and the effect on colour excess}
\label{sec:redden}
\begin{figure*}
\begin{center}
    \begin{tabular}{cc}
\resizebox{80mm}{!}{\includegraphics[angle=0]{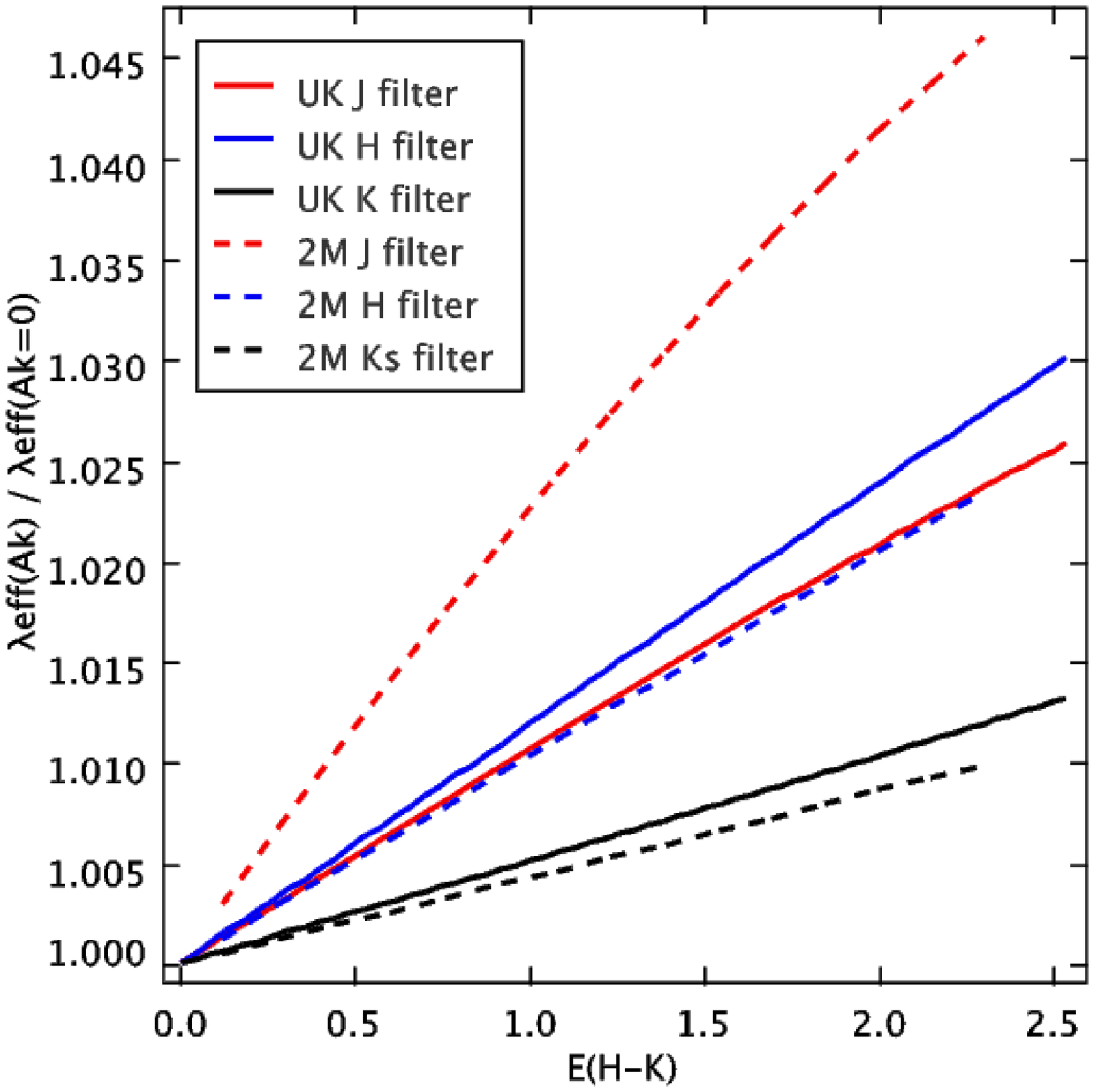}} &
\resizebox{80mm}{!}{\includegraphics[angle=0]{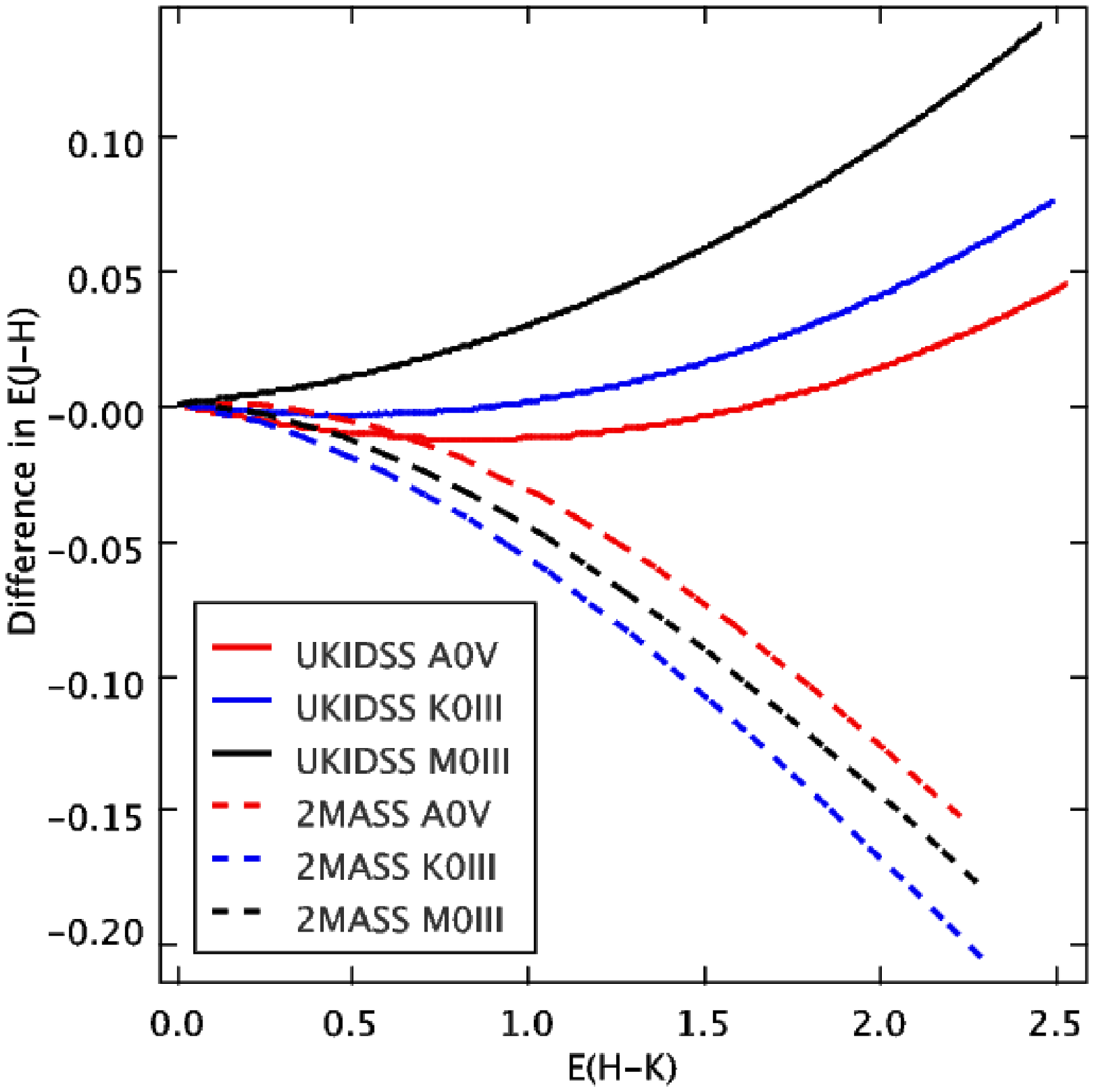}}
    \end{tabular}
\caption[]{\small  a (left): Illustrates how the effective wavelengths, $\lambda_{J_{eff}}$ (red lines), $\lambda_{H_{eff}}$ (blue lines) and $\lambda_{K_{eff}}$/$\lambda_{Ks_{eff}}$ (black lines) change as the spectra of an A0V CK04 model are progressively reddened. UKIDSS effective wavelengths are plotted as solid lines, 2MASS as dashed lines. b (right): Shows the reddening tracks of a progressively reddened (with a power law of $\alpha$=2.1) CK04 A0V model in red, a CK04 K0III model in blue and CK04 M0III model in black. Tracks created using the UKIDSS filters are shown as solid lines and those created using 2MASS filters are shown as dashed lines. Note the different directions of curvature.}
\label{fig:Vegaeff}
\end{center}
\end{figure*}

As mentioned previously in Section \ref{sec:intro}, it is commonly noted in the literature that effective filter wavelengths are colour dependent but the differences have usually been considered negligible. To investigate how the filter wavelengths change for sources with varying spectral type and A$_V$, spectra from the stellar atmosphere models by \citet{castelli04} (hereafter referred to as the CK04 stellar models) are reddened using a range of $\alpha$, with extinction in K (A$_K$) that varies between 0 and 3 mag in 0.005 intervals. The spectra are then convolved through the J, H and K UKIDSS \citep{hewett06} and J, H and K$_S$ 2MASS \citep{cohen03} filter profiles using the HST Synphot programme as described in \citet{hewett06}. \\
$\indent$By convolving stellar spectra through the filter profiles of each photometric system, we are able to model how the effective wavelengths depend upon source colour. It is this effect that causes the reddening $\textquoteleft \textquoteleft$vector$\textquotedblright$ to appear to curve as the spectra are progressively reddened. Fig.\ref{fig:Vegaeff}a illustrates the difference between each filter by comparing the ratio between the effective wavelength of each filter, created with the spectra of an A0V CK04 stellar model at zero reddening, and the effective wavelength of each filter as A$_K$ increases. \citet{golay74} defines the effective wavelengths of photometric filters as follows:
\begin{equation}
\lambda_{eff} = \frac{{\int}\lambda F({\lambda})S({\lambda}) d{\lambda}}{{\int} F({\lambda})S({\lambda}) d{\lambda}}.
  \label{eq:eff}
\end{equation}
$\indent$The significantly differing trends of the J filters, presented in Fig.\ref{fig:Vegaeff}a, cause the E$_{J-H}$ reddening track to reverse its direction of curvature between the two photometric systems, as shown in Fig.\ref{fig:Vegaeff}b. We use the term $\textquoteleft \textquoteleft$reddening track$\textquotedblright$ as opposed to the term $\textquoteleft \textquoteleft$reddening vector$\textquotedblright$ for two reasons. The term reddening vector is used to describe the magnitude and direction of the straight line along which a generic star moves, from its original position on a CCD, when observed through a specific amount of extinction. As the changing effective wavelengths cause this path to curve it can no longer be referred to as a vector. Also the amount of curvature this path possesses is dependent upon the spectral type of a given source. Throughout this paper the term $\textquoteleft \textquoteleft$reddening tracks$\textquotedblright$ is used to describe a group of tracks created with the same value of $\alpha$ but differing spectral type. The value of $\alpha$ used to create each reddening track is specified when necessary. Fig.\ref{fig:Vegaeff}b compares the difference in E$_{J-H}$, as E$_{H-K}$ increases, between the static (filter wavelengths at zero reddening for an A0V CK04 stellar model) and the evolving filter wavelengths of A0V, K0III and M0III CK04 stellar models as each are progressively reddened in A$_K$=0.005 intervals. Had the filter wavelengths not evolved with the progressively redder spectra, the reddening tracks created by each model, in either photometric system, would have remained at $\Delta$E$_{J-H}$=0. \\
$\indent$Using UKIDSS data, with typical errors of $\sim$0.04 in colour-colour space, it is not possible to resolve the differences between progressively redder spectra, or different reddening tracks, up to about E$_{H-K}$=1.2 (or A$_V$$\sim$15, using the reddening law discussed in Section \ref{sec:modelcdd-Aj}). It is therefore safe to assume static effective wavelengths for objects suffering from a low to intermediate amount of extinction. For heavily reddened objects, such as those found inside or behind star forming regions, this assumption of static effective wavelengths no longer holds. The progressively redder spectra in Fig.\ref{fig:Vegaeff}b display greatly differing amounts of curvature, and the curvature of each reddening track is not consistent between filters. For instance, when convolved through the UKIDSS filters, the M0III spectra posses the most curvature, however when convolved through the 2MASS filters it is the K0III spectra that possess the most curvature (in the opposite direction).
\section{Modelling the Colour Colour Diagrams}
\label{sec:modelcdd}
Each NIR extinction power law is derived by comparing data from UKIDSS and 2MASS, with a catalogue of synthetic stars created using the $\Bes$ stellar population synthesis model of the Galaxy.
\subsection{The Stellar Population Synthesis Model of the Galaxy}
\label{sec:modelcdd-Aj}
A complete description of the $\Bes$ model inputs can be found in \citet{robin03}, however it is summarised here for completeness. The model contains four populations of stars, thin disc, thick disc, spheroid and bulge. Each population is described by a star formation rate history, an initial mass function, an age or age range, a set of evolutionary tracks, kinematics, metallicity characteristics, and includes a white dwarf population. The extinction is modelled by a diffuse thin disc, described in terms of visual magnitudes per kiloparsec (mag/kpc.) The model also allows for the insertion of discrete clouds with a specified A$_V$ and distance. \\
$\indent$\citet{marshall06} use 2MASS and $\Bes$ model data to model the Galactic interstellar extinction distribution in three dimensions to over 64000 lines of sight, each separated by 15$^{\prime}$, in the inner Galaxy (hereafter referred to as the M06 distributions). In order to estimate the extinction along our chosen 8 lines of sight, we attempt to replicate the M06 distributions by using a set extinction model to describe the thin disc (mag/kpc) and insert discrete clouds at specified distances.\\ $\indent$The $\Bes$ model data contain the distance, the visual extinction, the apparent J, H and K magnitudes (given in the Johnson-Cousins system), and the spectral type (and therefore the intrinsic colours) of each synthetic star. The $\Bes$ intrinsic magnitudes are converted to the UKIDSS and 2MASS photometric systems and reddened to the desired extinction along the CK04 reddening tracks, using several different power laws (values of $\alpha$). \\
$\indent$To determine the extinction along each reddening track, the visual extinction (A$_V$) in the $\Bes$ model data is converted to an extinction in the J band (A$_J$) of each photometric system. For UKIDSS A$_J$/A$_V$=0.2833 and for 2MASS A$_J$/A$_V$=0.2900. These values are calculated using the \citet{cardelli89} extinction curve with R$_V$=3.1 (obtained from the Trilegal website http://stev.oapd.inaf.it/cgi-bin/trilegal). The value of A$_J$ for each synthetic star in the $\Bes$ model data is now matched with a value of A$_J$, and therefore the corresponding values of A$_H$ and A$_K$, in the desired reddening track. Using these values of A$_J$, A$_H$ and A$_K$ we can compute the apparent magnitudes of each synthetic star, in the desired photometric system, using a specific value of $\alpha$. The limiting magnitudes of the $\Bes$ model data are made consistent with those of the observed field and photometric system. This is beneficial since the photometric depths of each region can change depending on the chosen line of sight. As suggested in the previous section, stars of varying spectral type produce reddening tracks with differing amounts of curvature, particularly those created in the UKIDSS photometric system. This difference is noticeable at large amounts of reddening. For this reason, depending on the spectral type of each synthetic star, different reddening tracks created with CK04 stellar models of equivalent spectral type, are used to redden each star.
\subsection{The modelled regions}
\label{sec:longlat}
To determine accurate values of $\alpha$ we need to extract data from large areas to return an adequate number of stars. However the stellar population and therefore intrinsic colours may change over large areas. We have analysed the distribution of  the spectral types of giants along several lines of sight using $\Bes$ model data. From this we have found that the distribution of background giants can vary dramatically over a few tenths of a degree in Galactic latitude but remains constant over several degrees in Galactic longitude (with the exception of regions towards the Galactic centre). For this reason our chosen areas are typically l=60$^{\prime}$ x b=6$^{\prime}$ in size. As our method requires a large number of sources to determine a power law, in regions where the source count drops significantly (i.e. towards the outer edges of the Galaxy), we extend the area in the longitudinal direction to 90$^{\prime}$x6$^{\prime}$ in size. In order to avoid observing the Galactic bar, since this is not modelled in the $\Bes$ model data, we do not attempt regions within 27$^{\circ}$ of the Galactic centre \citep{garzon97}. Also, as there are no M06 distributions available with a Galactic longitude $>$100$^{\circ}$, we only consider regions between 27$^{\circ}$ and 100$^{\circ}$ in Galactic longitude.\\
\begin{figure}
  \begin{center}
    \begin{tabular}{cc}
      \resizebox{80mm}{!}{\includegraphics[angle=0]{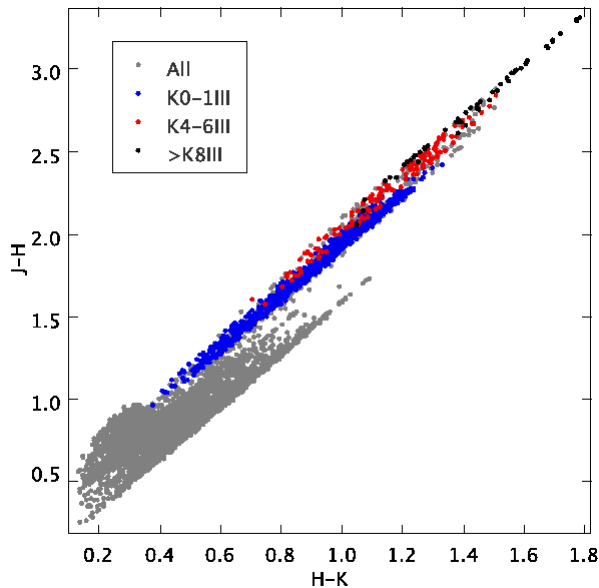}}
    \end{tabular}
    \caption{\small A CCD created from $\Bes$ model data that have not been converted from the Johnson-Cousins system and have therefore been reddened along a reddening vector. The data have been extracted from a region centred on the RMS source G48.9897-00.2992 (as in Fig.\ref{fig:exampleCCD}). There are 6075 sources and the apparent magnitude range is consistent with that of the UKIDSS data presented in  Fig.\ref{fig:exampleCCD}. The blue points cover stars with a spectral type range K0-K1III, the red K4-K6III and the black covers those with spectral type K8III and later. The giant branch appears to show some curvature due to the dominant observed spectral type varying with depth through the galactic plane.}
    \label{fig:popchange}
  \end{center}
\end{figure}
$\indent$Fig.\ref{fig:popchange} contains $\Bes$ model data extracted from the same region as the data in Fig.\ref{fig:exampleCCD}. As the dominant observed spectral type varies with depth through the galactic plane, so will the dominant intrinsic colours. This causes the giant locus to show some curvature even when reddened along a reddening vector. The separation of giant intrinsic colours is more pronounced in the UKIDSS photometric system than in the 2MASS photometric system. This causes the giant locus, as apparent in Fig.\ref{fig:exampleCCD}, to show more curvature when created using UKIDSS data than with 2MASS data.
\subsection{Comparison of model and observed CCDs}
\begin{figure*}
\begin{center}
    \begin{tabular}{ccc}
      \resizebox{55mm}{!}{\includegraphics[angle=0]{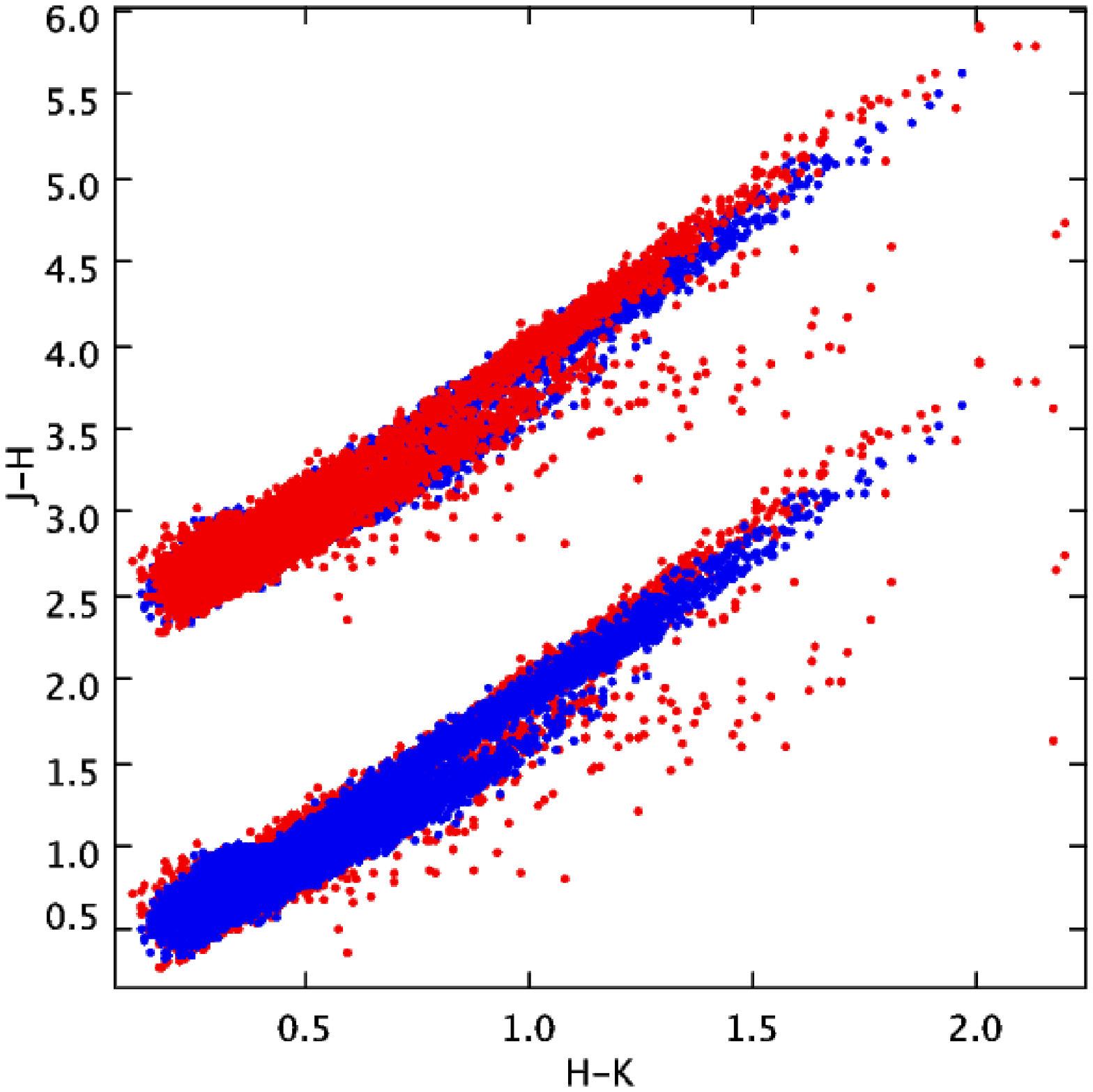}} &
\resizebox{55mm}{!}{\includegraphics[angle=0]{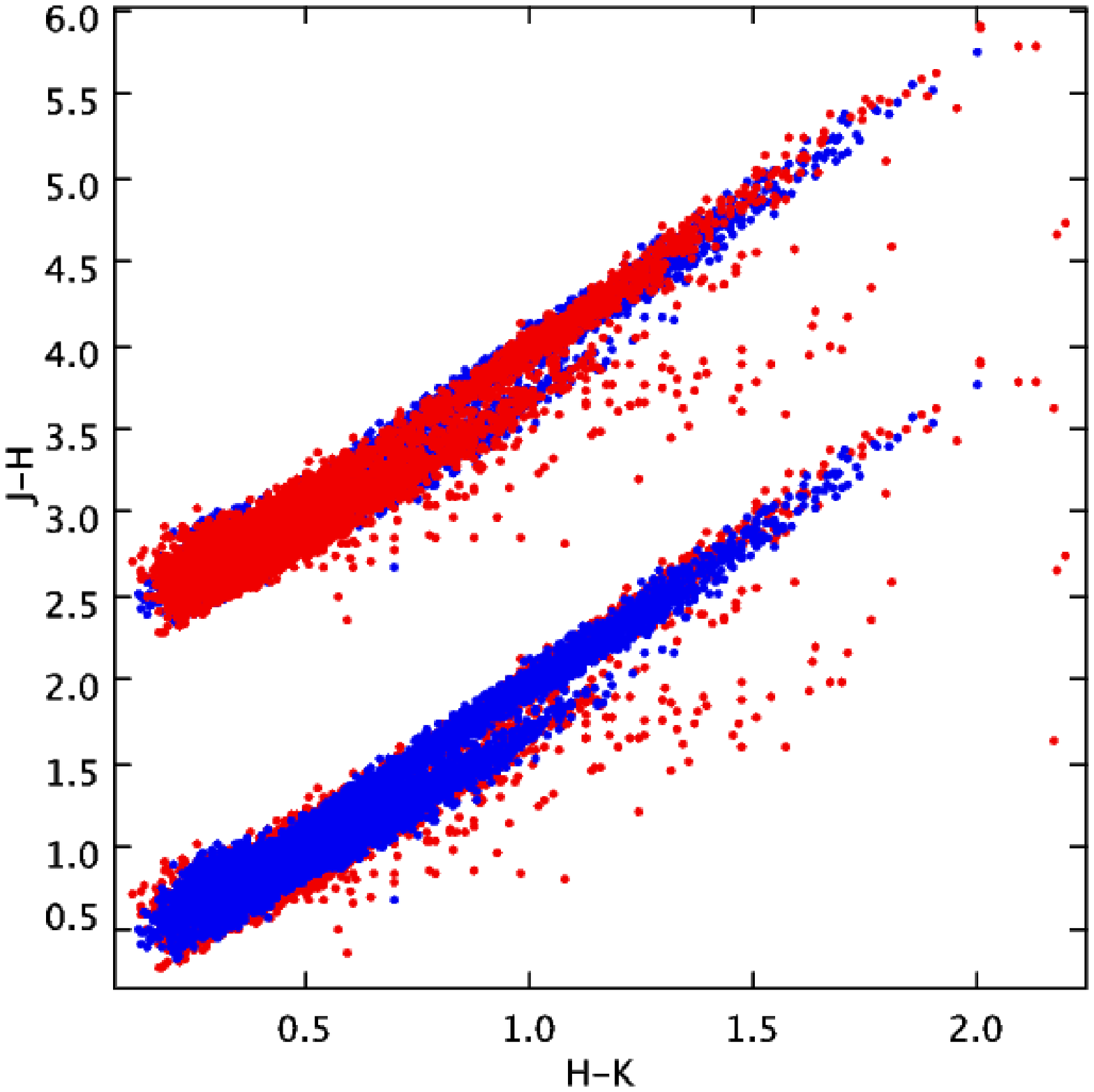}}&
\resizebox{55mm}{!}{\includegraphics[angle=0]{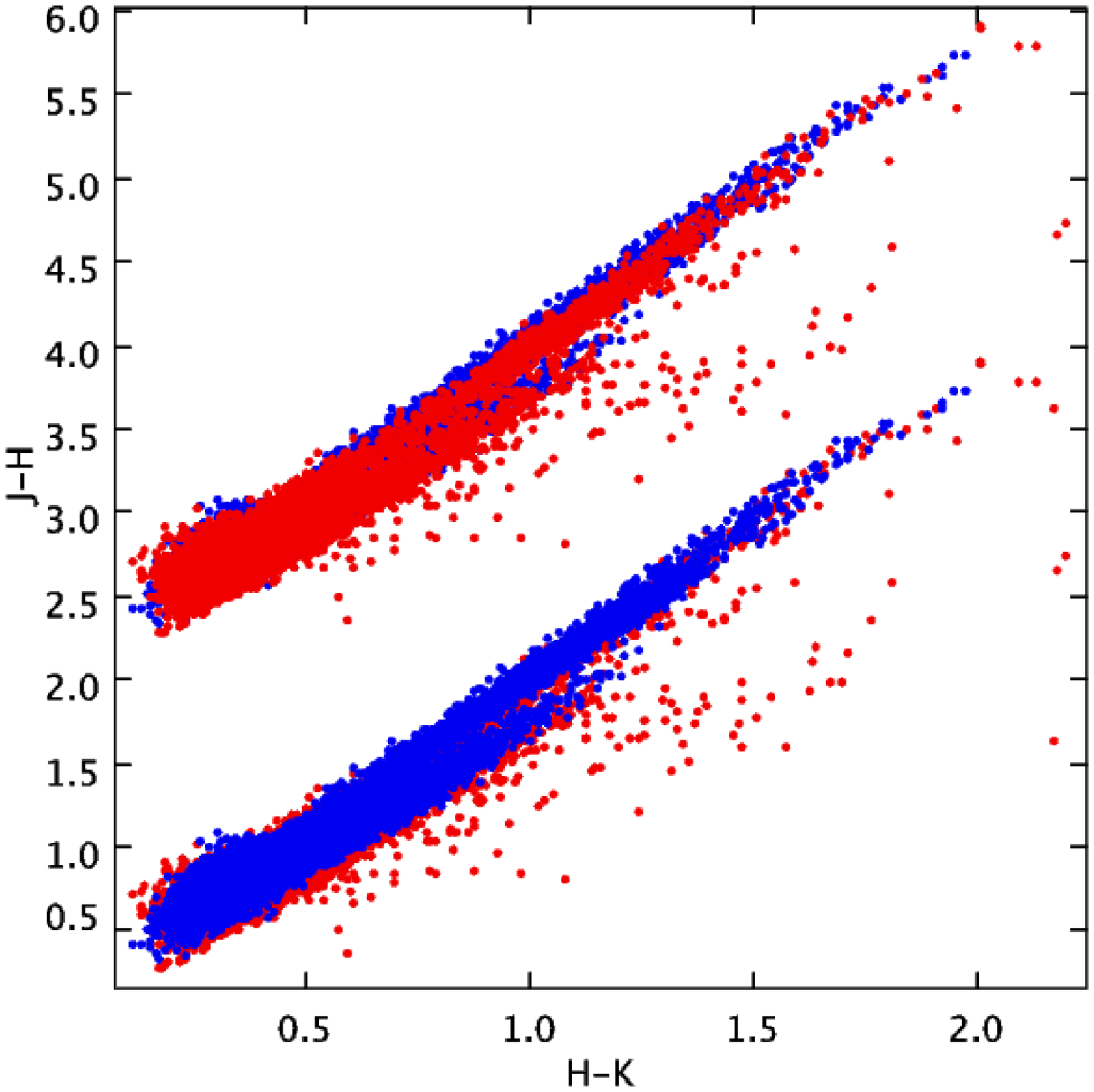}}
    \end{tabular} 
\caption[]{\small a (left): Contains UKIDSS data (red points) plotted against $\Bes$ model data (blue points), reddened along the CK04 reddening tracks created with $\alpha$=2.00. The UKIDSS and $\Bes$ model data are presented twice with an offset along the J-H axis and the overlay switched. b (centre): As (a) but with $\alpha$=2.15. c (right): As (a) but with $\alpha$=2.30. Unlike the UKIDSS data, the $\Bes$ model data do not contain any sources with an IR excess.}
\label{fig:ukidssfit}
\end{center}
\end{figure*}

The $\Bes$ model data are reddened using a range of power laws and compared to real data in colour-colour space. To determine which power law creates the best fit between the synthetic and real data we aim to minimise the following quantity:
\begin{equation}
L =\frac{1}{n}\sum^{n}_{i=1}((S_{(H-K)_{i}}-R_{(H-K)_{i}})^{2}+(S_{(J-H)_{i}}-R_{(J-H)_{i}})^{2})^{\frac{1}{2}},
  \label{eq:tau}
\end{equation}
where L is defined as the mean separation, in colour-colour space, between each real star (R) and its nearest synthetic star neighbour (S). We perform Monte Carlo simulations to calculate the 1$\sigma$ deviation of L and from this a value of $\alpha$, with 1$\sigma$ errors, is determined. We also sample from a realistic distribution of photometric errors in the simulation to ensure we are treating the model data in as similar a way as possible to the real data. Typical observational photometric errors are created using an exponential function, that describes the relationship between apparent magnitude and photometric error, (e.g. Jerr = A + exp(B x Jmag - C), where A, B and C are coefficients specific to each photometric system) and a random Gaussian distribution. Unlike the real data, the synthetic data do not contain sources with an IR excess. To avoid any contamination in the fitting procedure, potential IR excess sources are removed from the real sample before a fit is made. This process is done by eye by removing any sources that extend away from the bulk of the data, as illustrated in Fig.\ref{fig:exampleCCD}. In each of the 8 regions such stars account for $\sim$1$\%$ of total sources and therefore have little effect on the fit. \\
$\indent$As an example we use data, extracted from around the RMS source G48.9897-00.2992, shown in Fig.\ref{fig:exampleCCD}. G48.9897 has been identified as a HII region with a derived kinematic distance of 5.2kpc. Initially we first derive extinction laws using $\Bes$ model data created with a simple extinction distribution models with a fixed amount of extinction per distance (mag/kpc). To determine the deviation of L, we compare 200 different sets of $\Bes$ model data with the real data from the surrounding area. So we require 200 times as many synthetic stars as there are real stars (in this particular case there are $\sim$6,640 UKIDSS sources and so $\sim$1,330,000 synthetic stars are required) to draw 200 different random selections of stars, each containing a similar number of stars as the real Galactic field, in order to compare with the real data. A large amount of $\Bes$ model data are extracted from a pencil beam centred on the RMS source, thereby effectively giving every synthetic star the same l and b as the RMS source. As mentioned in Section \ref{sec:longlat}, we avoid the problem of the changing stellar population by restricting the range of Galactic latitudes used to extract the UKIDSS and 2MASS data. \\
\begin{figure*}
\begin{center}
    \begin{tabular}{ccc}
      \resizebox{51mm}{!}{\includegraphics[angle=0]{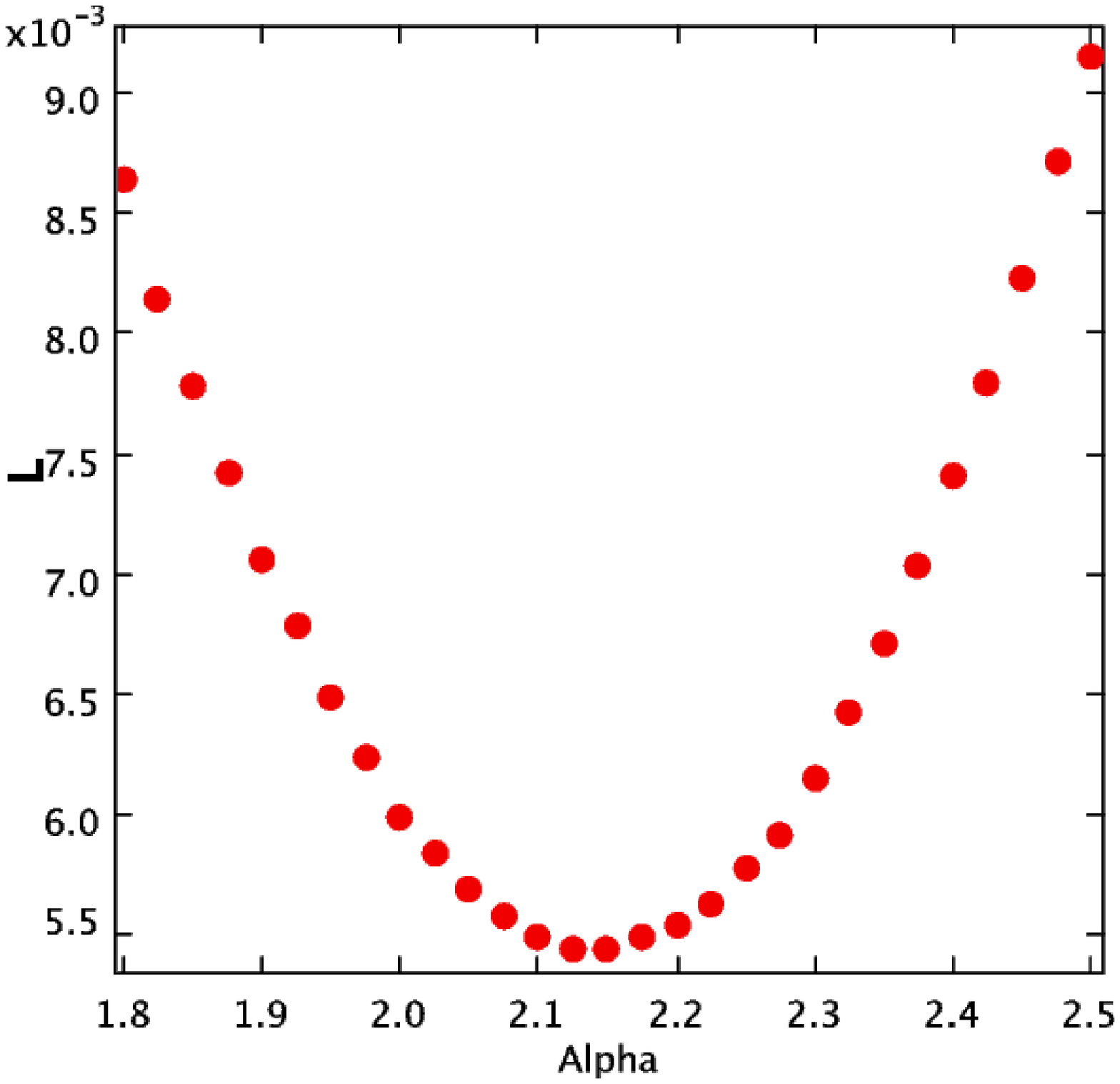}} &
\resizebox{65mm}{!}{\includegraphics[angle=90]{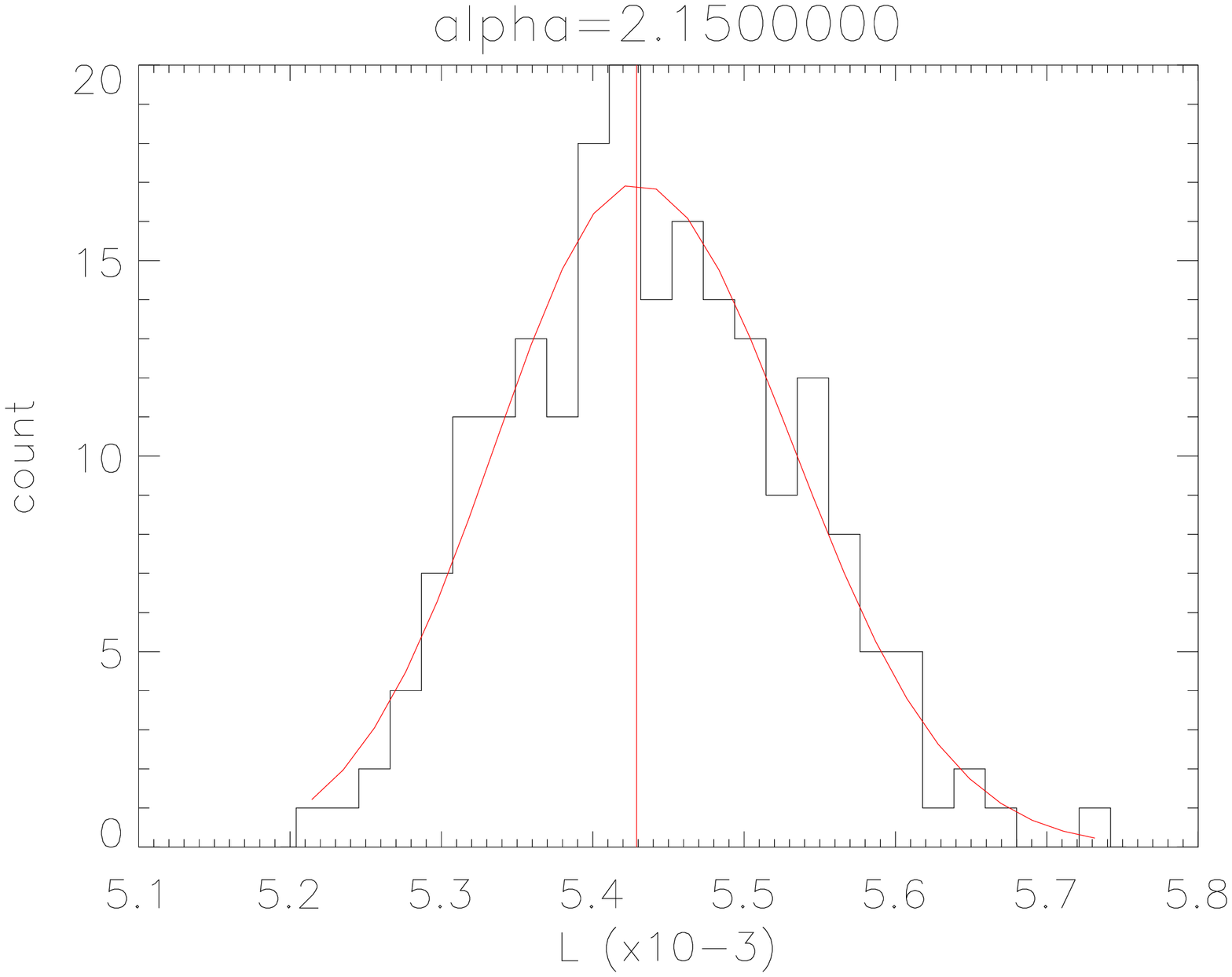}}&
\resizebox{51mm}{!}{\includegraphics[angle=0]{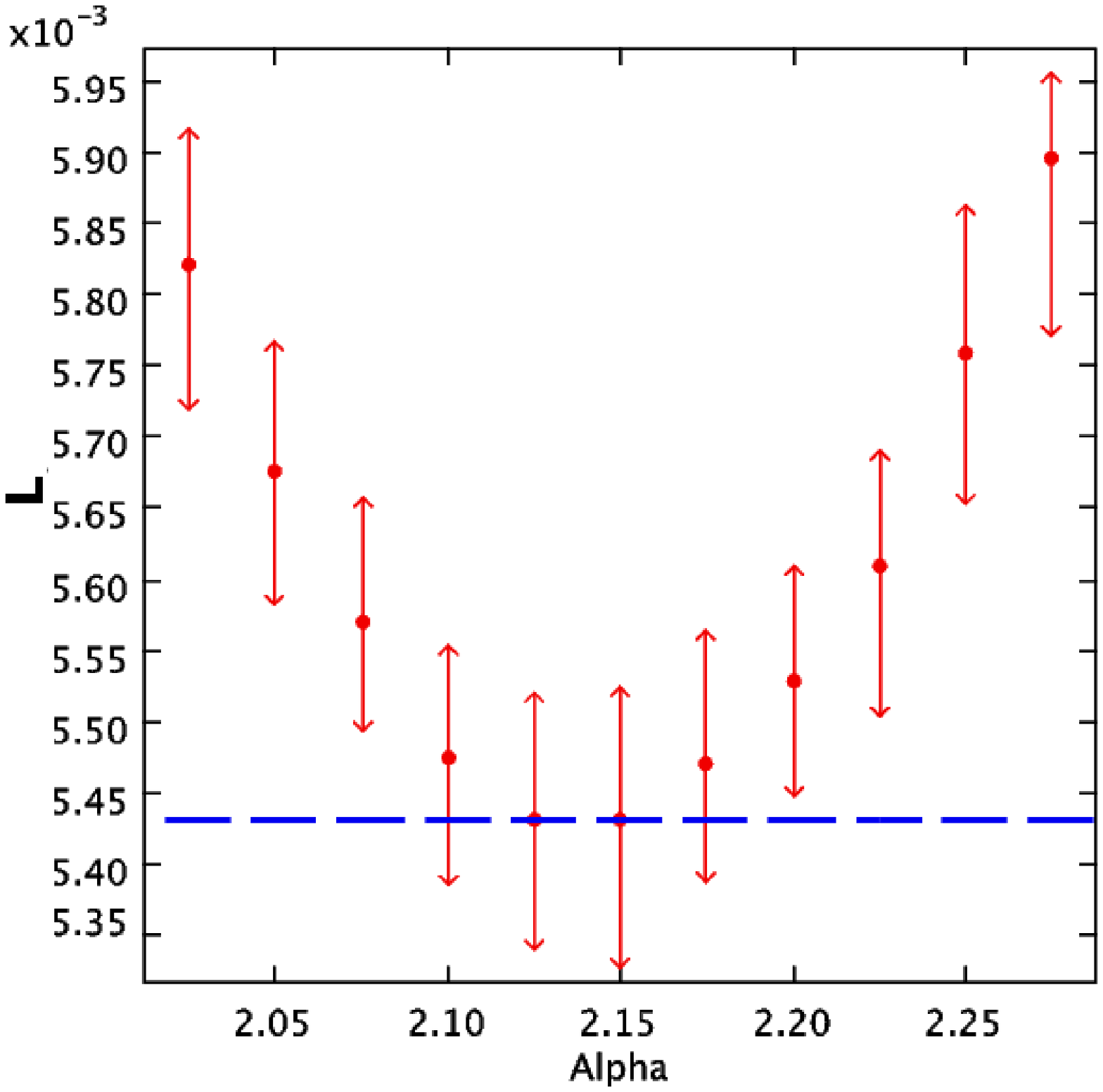}}
    \end{tabular} 
\caption[]{\small a (left): $\alpha$ vs L using one randomly drawn set of $\Bes$ model data. The smallest value of L occurs when the $\Bes$ model data best fit the real data. b (centre): The distribution of L for 200 different sets of $\Bes$ model data when $\alpha$=2.15. Also plotted are the fitted Gaussian and the median. c (right): The values of L where the 1$\sigma$ deviation error bar pass below the dashed blue line (smallest median) define the random error in $\alpha$.}
\label{fig:chi_compare}
\end{center}
\end{figure*}
$\indent$Fig.\ref{fig:chi_compare}a illustrates how L varies with $\alpha$ using one randomly drawn set of $\Bes$ model data created with an extinction distribution model of 1.5mag/kpc. The smallest value of L occurs when the $\Bes$ model data best fit the real data, and in this particular case this occurs when $\alpha$=2.150. This process is repeated 200 times to determine the distribution of L for each value of $\alpha$. Fig.\ref{fig:chi_compare}b contains a histogram illustrating this distribution, with a skewed Gaussian fitted to determine the median and 1$\sigma$ deviation. The value of $\alpha$ that creates an L distribution with the smallest median defines the final value of $\alpha$. The random error is defined by the values of $\alpha$ where the 1$\sigma$ deviation subtracted from the median value of L, is smaller than the smallest median. This process is illustrated in Fig.\ref{fig:chi_compare}c where the values of $\alpha$=2.100 to 2.175 are within a 1$\sigma$ deviation of the smallest median. \\
$\indent$The fitting process is illustrated using $\Bes$ model data, reddened using three different extinction laws along the CK04 reddening tracks, and UKIDSS data in Fig.\ref{fig:ukidssfit}. The final value for UKIDSS data, using an extinction distribution model of 1.5mag/kpc, is $\alpha$=2.150$\pm ^{0.060}_{0.084}$, these errors include accounting for the discretisation of $\alpha$. The process is repeated using 2MASS data deriving a final value of $\alpha$=2.025$\pm ^{0.181}_{0.081}$. These two values of $\alpha$ agree within 1$\sigma$ uncertainties and give an average law of $\alpha$=2.088$\pm ^{0.095}_{0.059}$. This value of $\alpha$ is significantly larger than the majority of values present in the literature. All derived extinction laws are presented in Table \ref{tabz:power}. 

\subsection{The extinction distribution}
\label{sec:extinctiondistribution}
\begin{figure*}
\begin{center}
    \begin{tabular}{ccc}
      \resizebox{70mm}{!}{\includegraphics[angle=90]{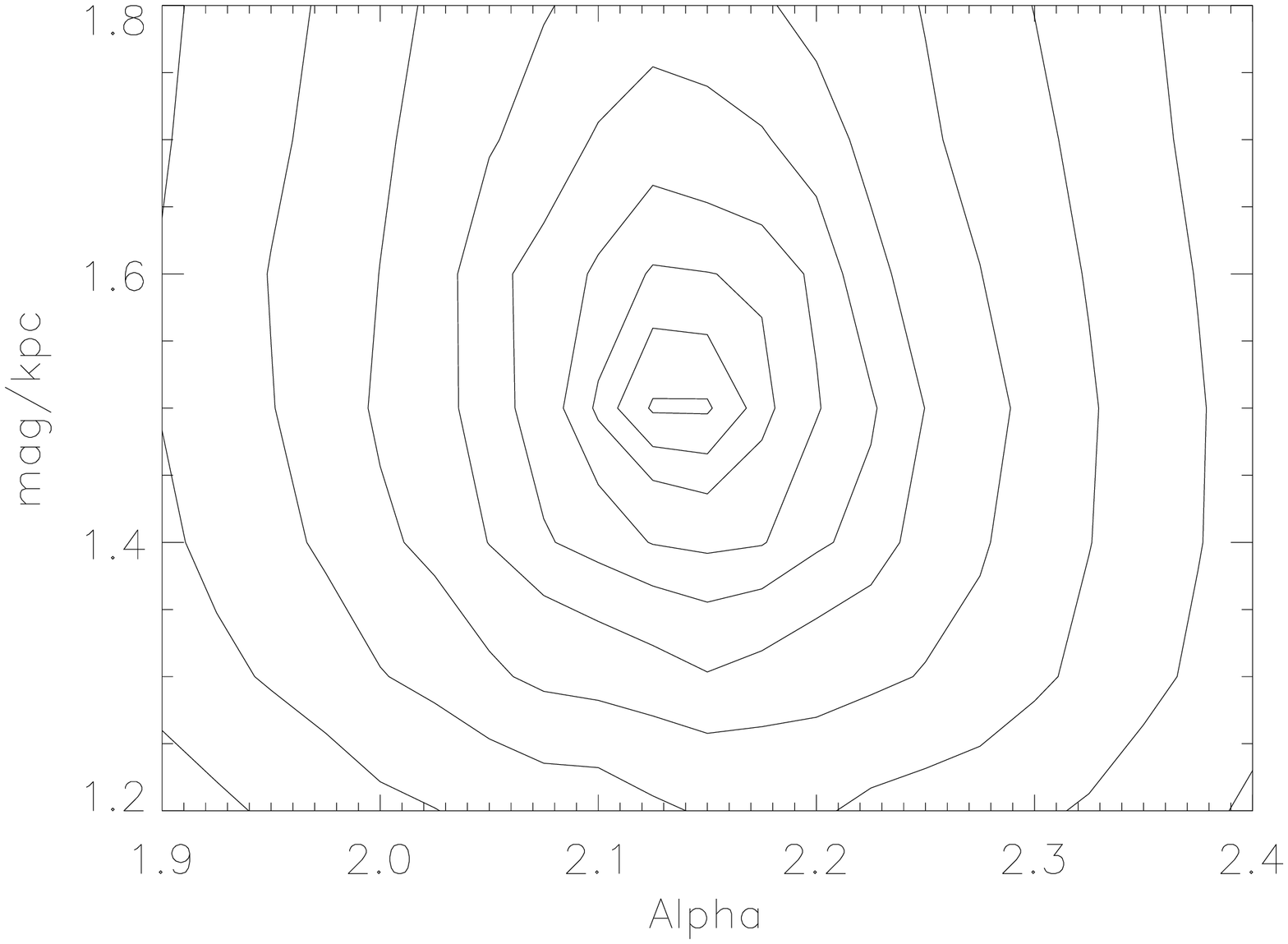}} &
\resizebox{55mm}{!}{\includegraphics[angle=0]{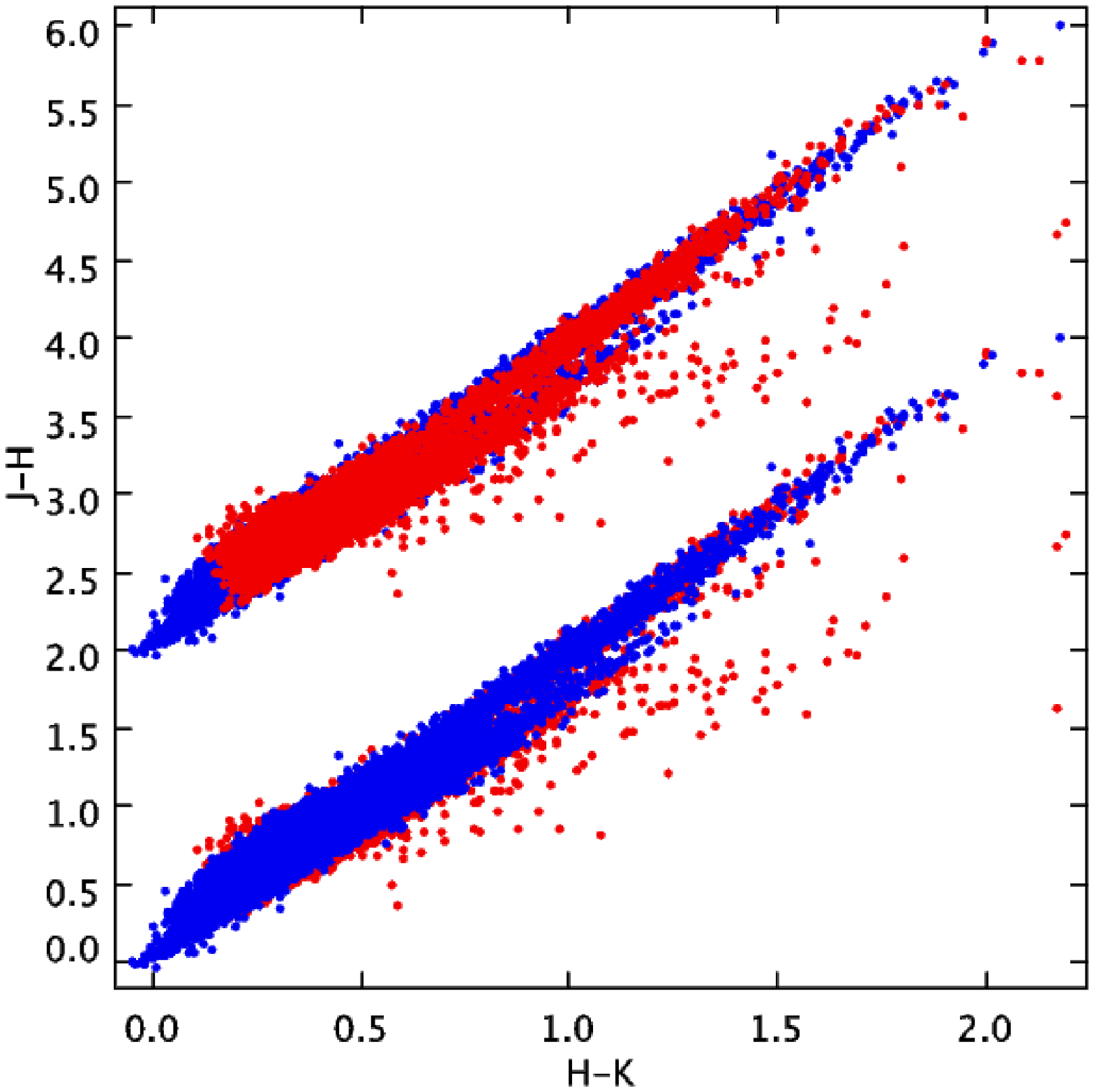}}
    \end{tabular} 
\caption[]{\small a (left): Simultaneous minimisation of L as the extinction model (mag/kpc) and value of $\alpha$ are varied. b (right): Contains UKIDSS data (red points) plotted against $\Bes$ model data (blue points) created by randomising the visual extinction each source suffers (see text) and reddened along the CK04 reddening tracks created with $\alpha$=2.225.}
\label{fig:varying_extinction}
\end{center}
\end{figure*}
\begin{figure*}
\begin{center}
    \begin{tabular}{ccc}
\resizebox{60mm}{!}{\includegraphics[angle=0]{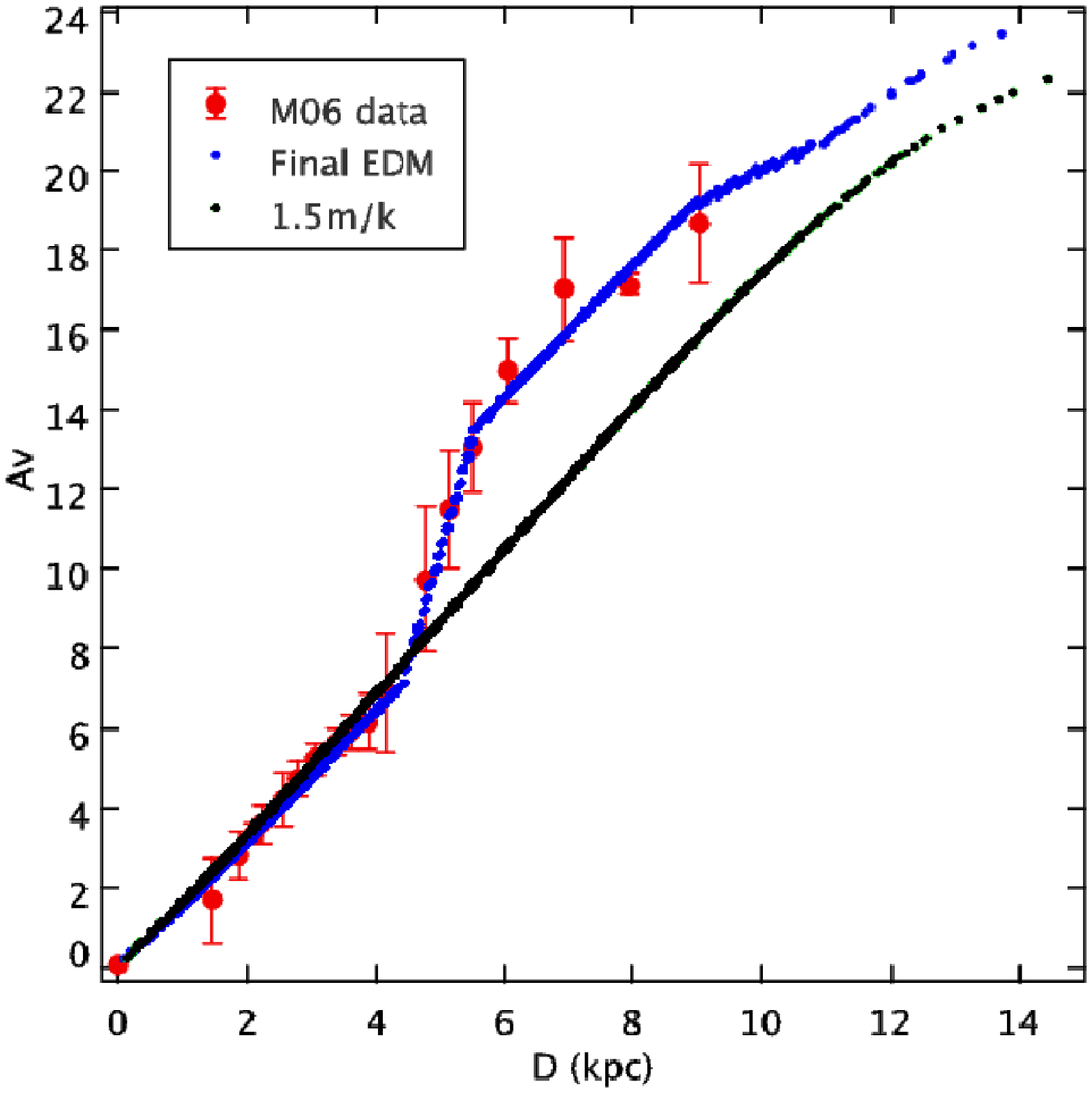}}&
      \resizebox{60mm}{!}{\includegraphics[angle=0]{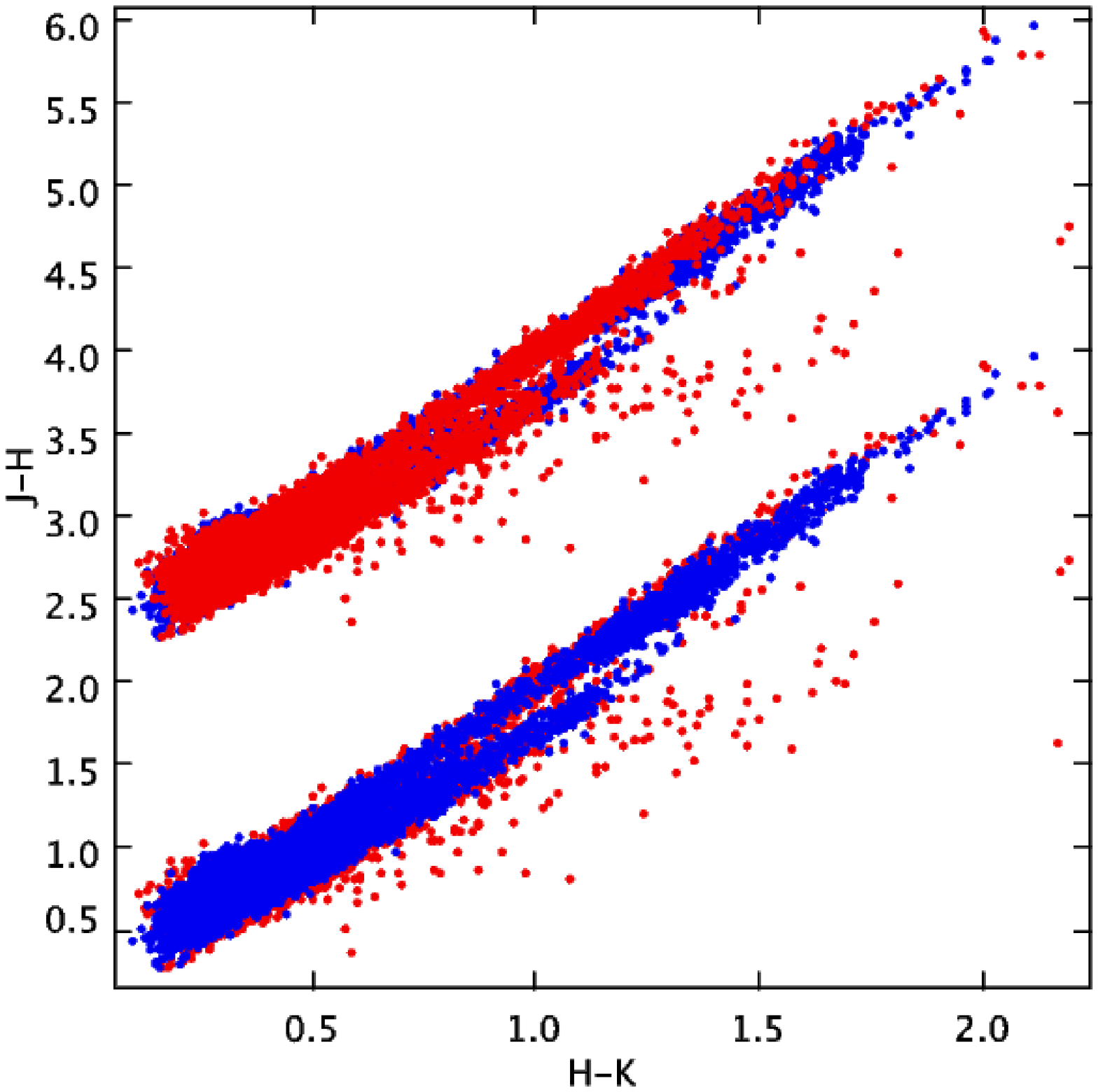}} &
\resizebox{60mm}{!}{\includegraphics[angle=0]{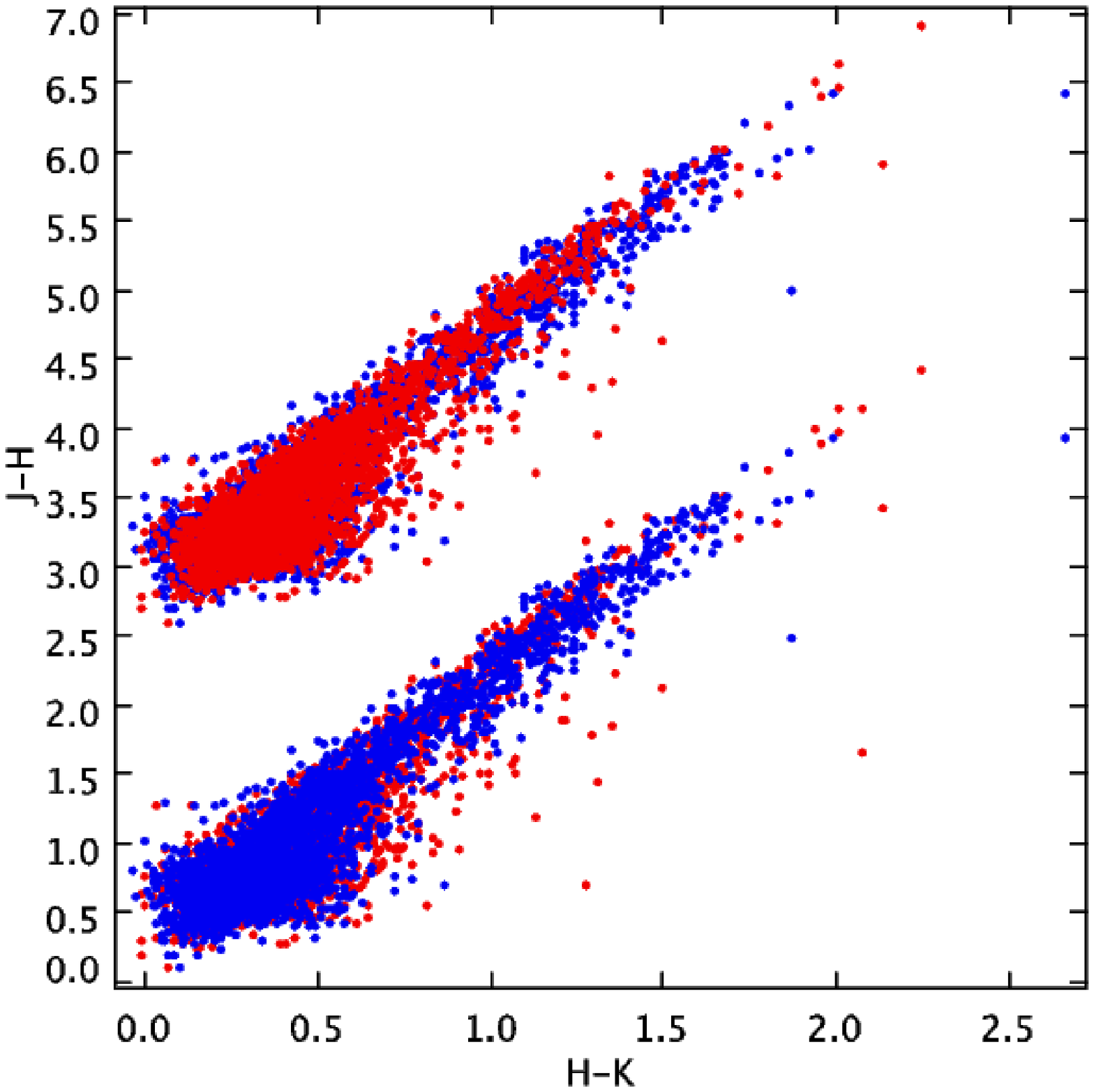}}
    \end{tabular} 
\caption[]{\small a (right): A distance-A$_V$ plot containing the extinction distribution model, lying within the constraints of the M06 data, that is used to derive each extinction power law for the region G48.9897. b (centre): A CCD created using UKIDSS (red points) and $\Bes$ data (blue points) reddened with the derived law, $\alpha$=2.15. c (left): As (c) but with 2MASS data and $\alpha$=2.05.}
\label{fig:g48.9897}
\end{center}
\end{figure*}

To test how dependent the derived extinction power law is on the extinction distribution model used to create the $\Bes$ model data, we perform two tests. First we vary the simple extinction distribution model of 1.5mag/kpc over a range of intervals between 1.2mag/kpc and 1.8mag/kpc corresponding to a change of $\pm$20$\%$. Fig.\ref{fig:varying_extinction}a contains a contour plot illustrating the simultaneous minimisation of L as the extinction model (mag/kpc) and value of $\alpha$ vary. The elongated shape in the direction of the y-axis suggests that the fitting technique is not strongly affected by the change in extinction per kpc. Table \ref{tabz:power} contains the laws derived using the 1.2mag/kpc and 1.8mag/kpc extinction distribution models, in both photometric systems. Each is consistent to within 1$\sigma$ errors of the laws derived using the 1.5mag/kpc extinction distribution model. \\
$\indent$For the second test we randomise the amount of extinction each source suffers in order to represent the patchiness of extinction over the observed region. The maximum amount of extinction suffered by any source in the $\Bes$ model data for the region G48.9897, reddened with an extinction model of 1.5mag/kpc, is A$_V$$\sim$20. An amount of extinction, randomly selected from a Gaussian distribution with mean A$_V$=0 and $\sigma$=10mag, was added to (or subtracted from) each star. In such cases where the subtraction of a random amount of extinction would result in A$_V$$<$0, the original value of A$_V$ in the $\Bes$ model data was used. This test has produced some realistic features of a CCD and some unrealistic features. The CCD now contains a small number of sources that have extended away from the bulk of the data with H-K$>$1.8 (Fig.\ref{fig:varying_extinction}b) similar to the real UKIDSS data. However, now there is also an over abundance of sources with a very low A$_V$. This creates the blue spike in Fig.\ref{fig:varying_extinction}b that appears between H-K $\sim$-0.1 and $\sim$0.1. Although this feature is unrealistic, the small amount of extinction suffered by these sources means their presence will have little effect on the fit. Due to the scatter created by these randomised data, the error in the derived extinction power law is noticeably larger than the law derived using an extinction model of 1.5mag/kpc and UKIDSS data. Due to this extra scatter, it was not possible to constrain a value using 2MASS data. Table \ref{tabz:power} contains the derived extinction power law and it is consistent with the extinction power law derived using the 1.5mag/kpc extinction model. \\
$\indent$For the region G48.9897, the UKIDSS and 2MASS data generate CCDs that appear smooth and somewhat uniform, being that there do not appear to be any large clumps or gaps in the data. It is as though all sources along the line of sight roughly suffer from a constant amount of extinction per kpc. This is why the $\Bes$ model data created with a simple extinction distribution model, of 1.5mag/kpc, produces such a realistic looking CCD. Data extracted from many of our other regions often do not produce such smooth and uniform CCDs. This is because the molecular cloud surrounding each star forming region analysed often creates a sudden jump in extinction, affecting stars inside and behind the cloud producing CCDs with gaps in the data. For this reason it is not always possible to assume sources in the field suffer from a constant amount of extinction per kpc. In such cases a more sophisticated extinction distribution model will be needed to detail how A$_V$ varies with distance. As discussed in section \ref{sec:modelcdd}, we use the M06 distributions to describe the extinction along the line of sight. We use the M06 distribution that is spatially closest to the RMS source to attempt to match the detailed morphology of the CCDs created using UKIDSS and 2MASS data. Each of the M06 distributions are replicated by using a set extinction model to describe the thin disc (mag/kpc) and discrete clouds inserted at specific distances. \\
$\indent$A distance vs A$_V$ plot for the region G48.9897 is presented in Fig.\ref{fig:g48.9897}a containing the M06 data and extinction distribution model used to derive extinction power laws, consistent with the laws derived using the  1.5mag/kpc extinction model. The new extinction distribution model possesses a sharp rise in A$_V$ at $\sim$5kpc; this is consistent with a massive star forming region situated at the RMS kinematic distance of 5.2kpc along the line of sight. There is only a small improvement in the size of L which is likely to be why the exact same value of $\alpha$ results in the best fit for the UKIDSS data. The size of the error however is reduced slightly. Table \ref{tabz:power} contains the derived extinction power laws for each photometric system.\\
$\indent$The 1.5mag/kpc extinction model is also plotted in Fig.\ref{fig:g48.9897}a. Despite the constant amount of extinction suffered per kpc, the 1.5mag/kpc model (black line) begins to curve after $\sim$12kpc. This effect is present (to a greater or lesser extent) in the majority of the M06 models used in this paper. It is an output of the $\Bes$ model and occurs as the edge of the synthetic Galaxy, and therefore the end of the thin disk, is approached. Fig.\ref{fig:g48.9897}b-c display the resultant CCDs using UKIDSS and 2MASS data, created using the final extinction distribution model. Note a slight deficit of model data (blue points) in Fig.\ref{fig:g48.9897}b around H-K$\sim$1.0, in comparison to Fig.\ref{fig:ukidssfit}b, due to the increase in the extinction distribution. \\
$\indent$To test the dependence of the results on the assumed intrinsic colours, a second set of intrinsic colours and reddening tracks were created from the stellar spectral flux library by \citet{pickles98} (hereafter referred to as P98). The process has been repeated, using all the extinction distribution models detailed in this paper, with the P98 intrinsic colours and reddening tracks. The results derived for each of the extinction distribution models, using both UKIDSS and 2MASS data, were consistent with results derived using the CK04 intrinsic colours and reddening tracks.
\section{Results}
\label{sec:results}
\begin{figure*}
\begin{center}
    \begin{tabular}{cc}
      \resizebox{60mm}{!}{\includegraphics[angle=0]{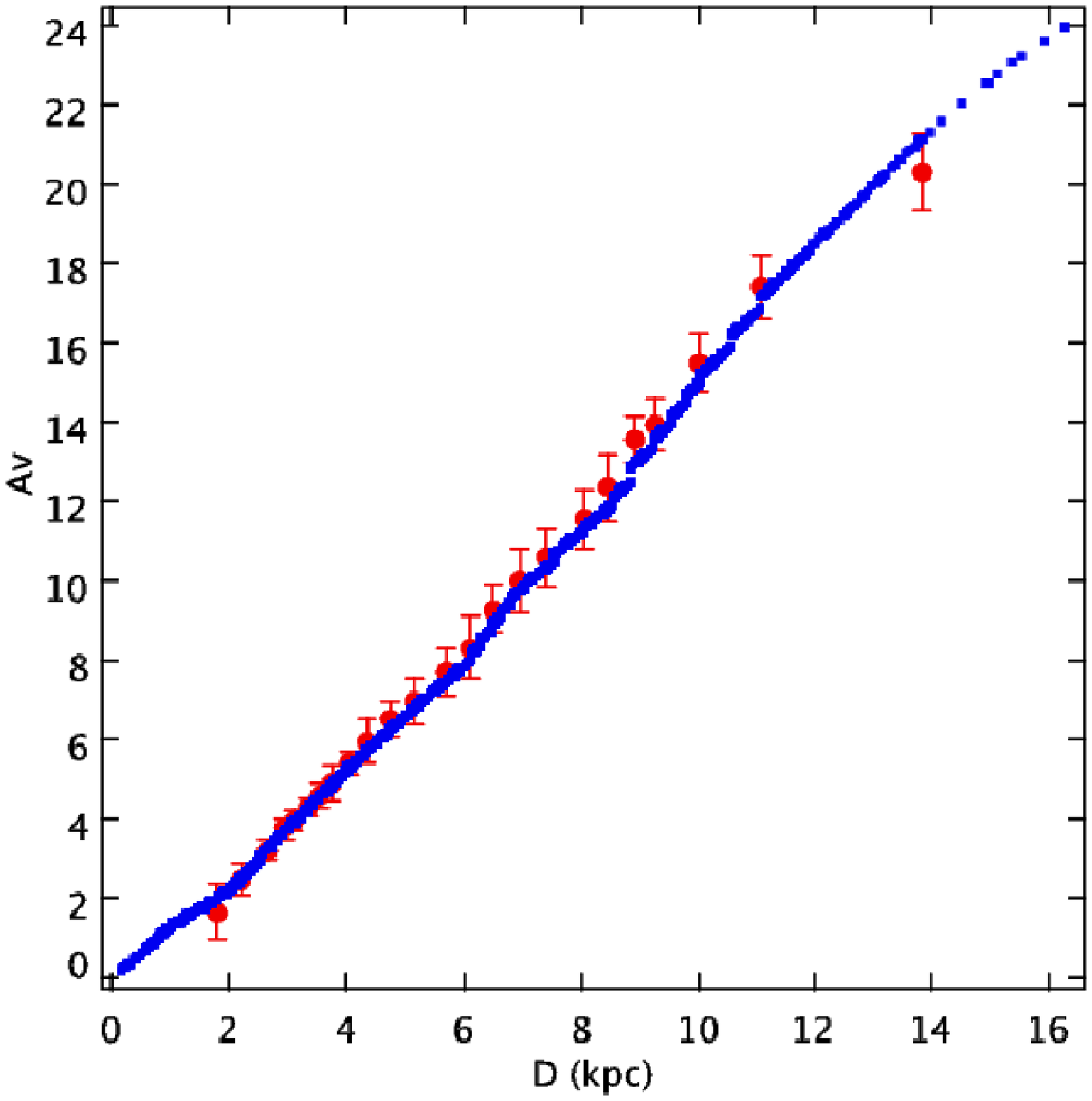}} &
\resizebox{60mm}{!}{\includegraphics[angle=0]{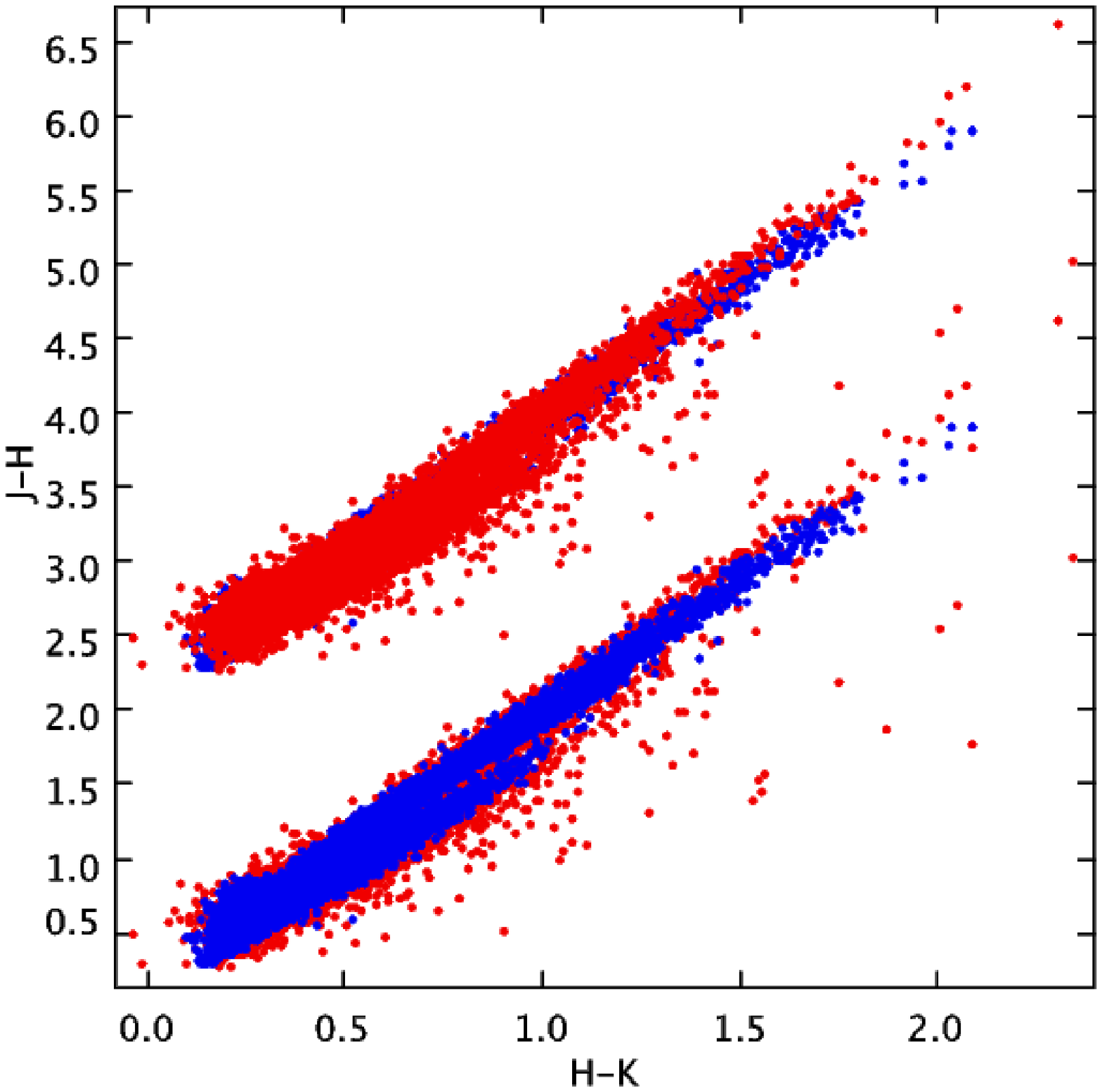}}\\
\resizebox{60mm}{!}{\includegraphics[angle=0]{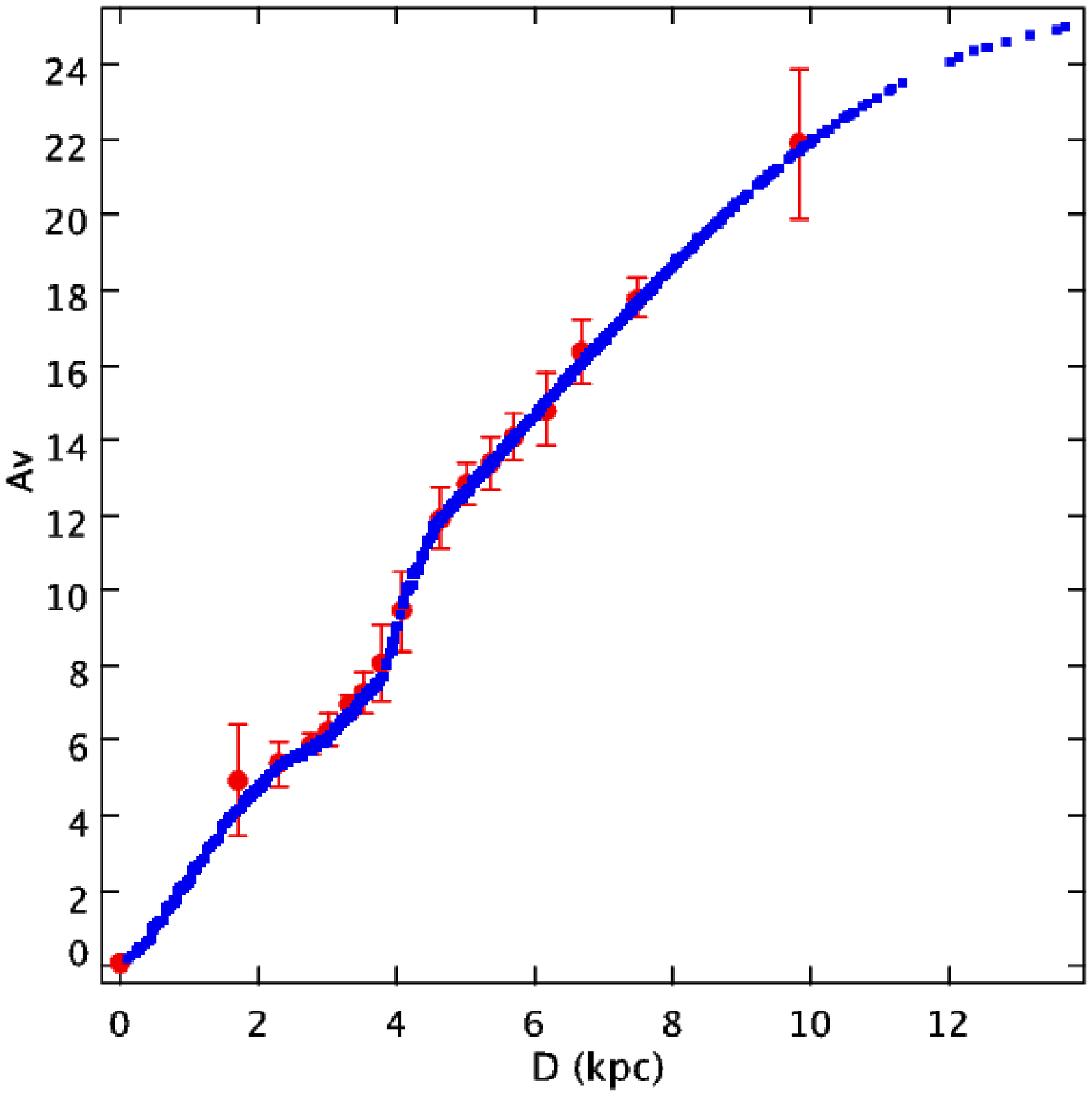}}&
      \resizebox{60mm}{!}{\includegraphics[angle=0]{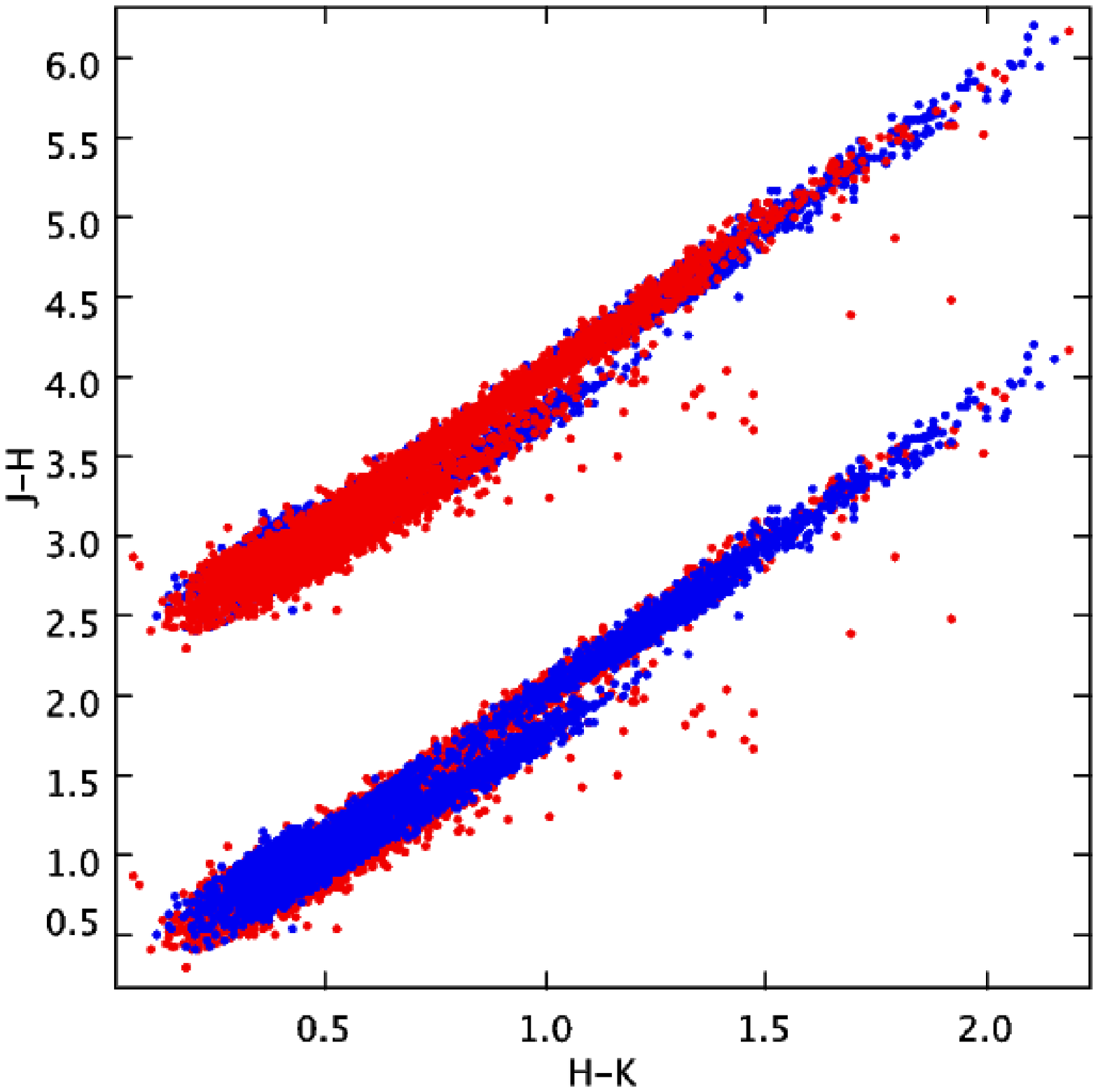}} \\
\resizebox{60mm}{!}{\includegraphics[angle=0]{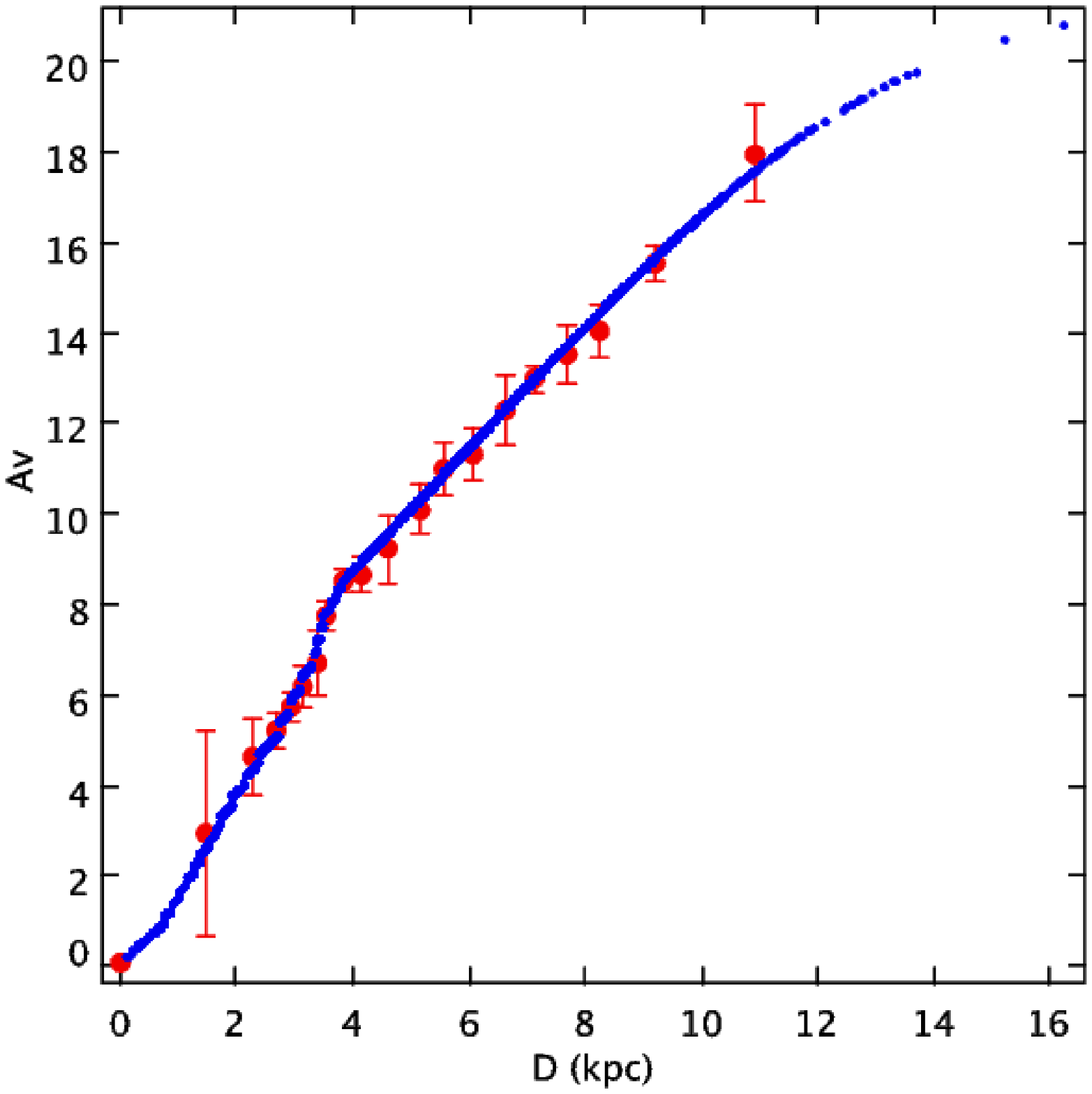}}&
      \resizebox{54mm}{!}{\includegraphics[angle=0]{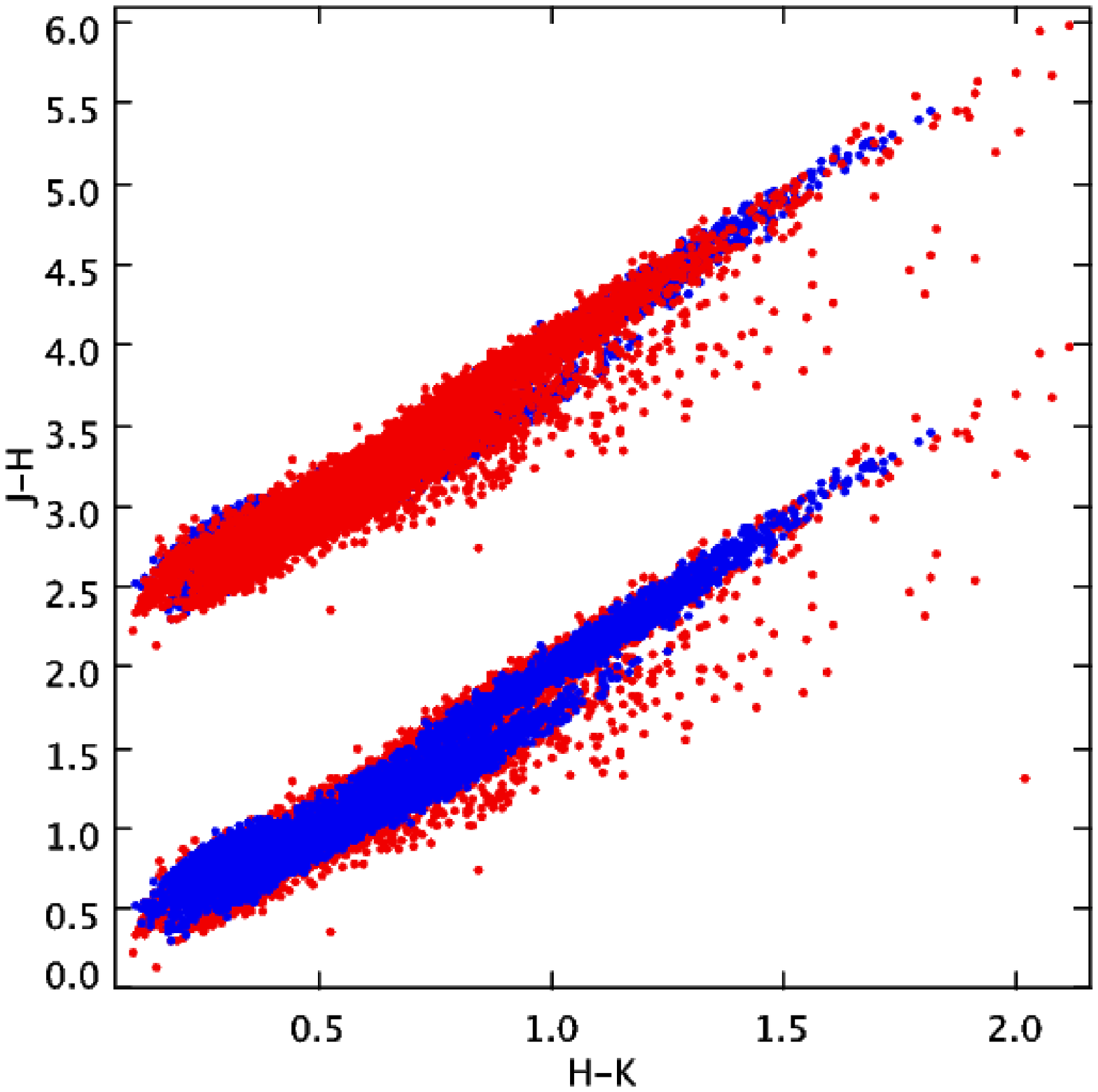}} 
    \end{tabular} 
\caption[]{\small a (top left): A distance-A$_V$ plot containing the extinction distribution model, lying within the constraints of the M06 data, that is used to derive each extinction power law for the region G29.9564. b (top right): A CCD created using UKIDSS data extracted from the region G29.9564 (red points) and $\Bes$ data (blue points) reddened with the derived law, $\alpha$=2.20. c (centre left): A distance-A$_V$ plot containing the extinction distribution model, lying within the constraints of the M06 data, that is used to derive each extinction power law for the region G42.1268. d (centre right): A CCD created using UKIDSS data extracted from the region G42.1268 (red points) and $\Bes$ data (blue points) reddened with the derived law, $\alpha$=2.30. e (bottom left): A distance-A$_V$ plot containing the extinction distribution model, lying within the constraints of the M06 data, that is used to derive each extinction power law for the region G53.6185. f (bottom right): A CCD created using UKIDSS data extracted from the region G53.6185 (red points) and $\Bes$ data (blue points) reddened with the derived law, $\alpha$=2.175.}
\label{fig:g29.9564}
\end{center}
\end{figure*}

\begin{figure*}
\begin{center}
    \begin{tabular}{cc}
\resizebox{60mm}{!}{\includegraphics[angle=0]{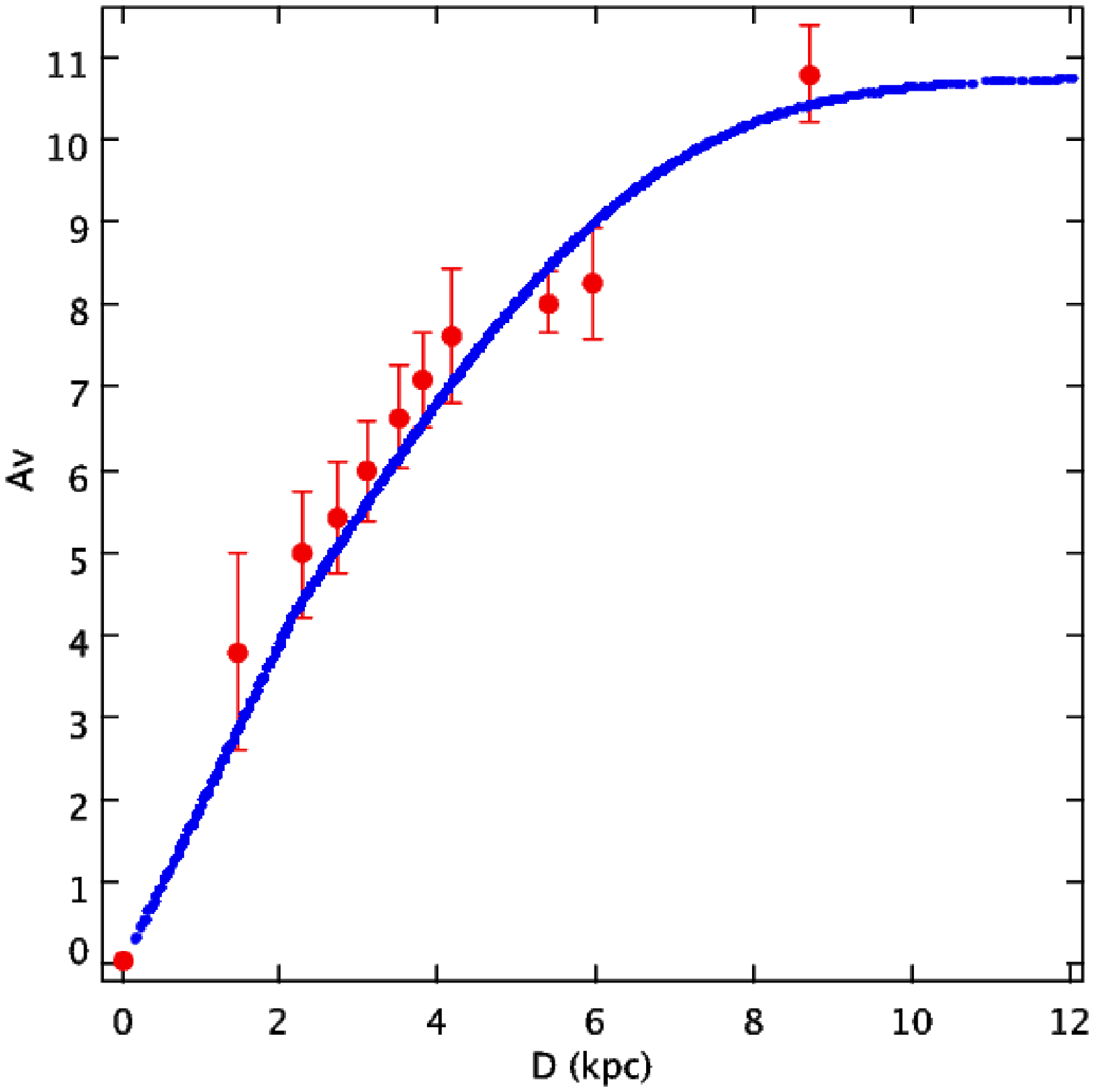}} &
      \resizebox{60mm}{!}{\includegraphics[angle=0]{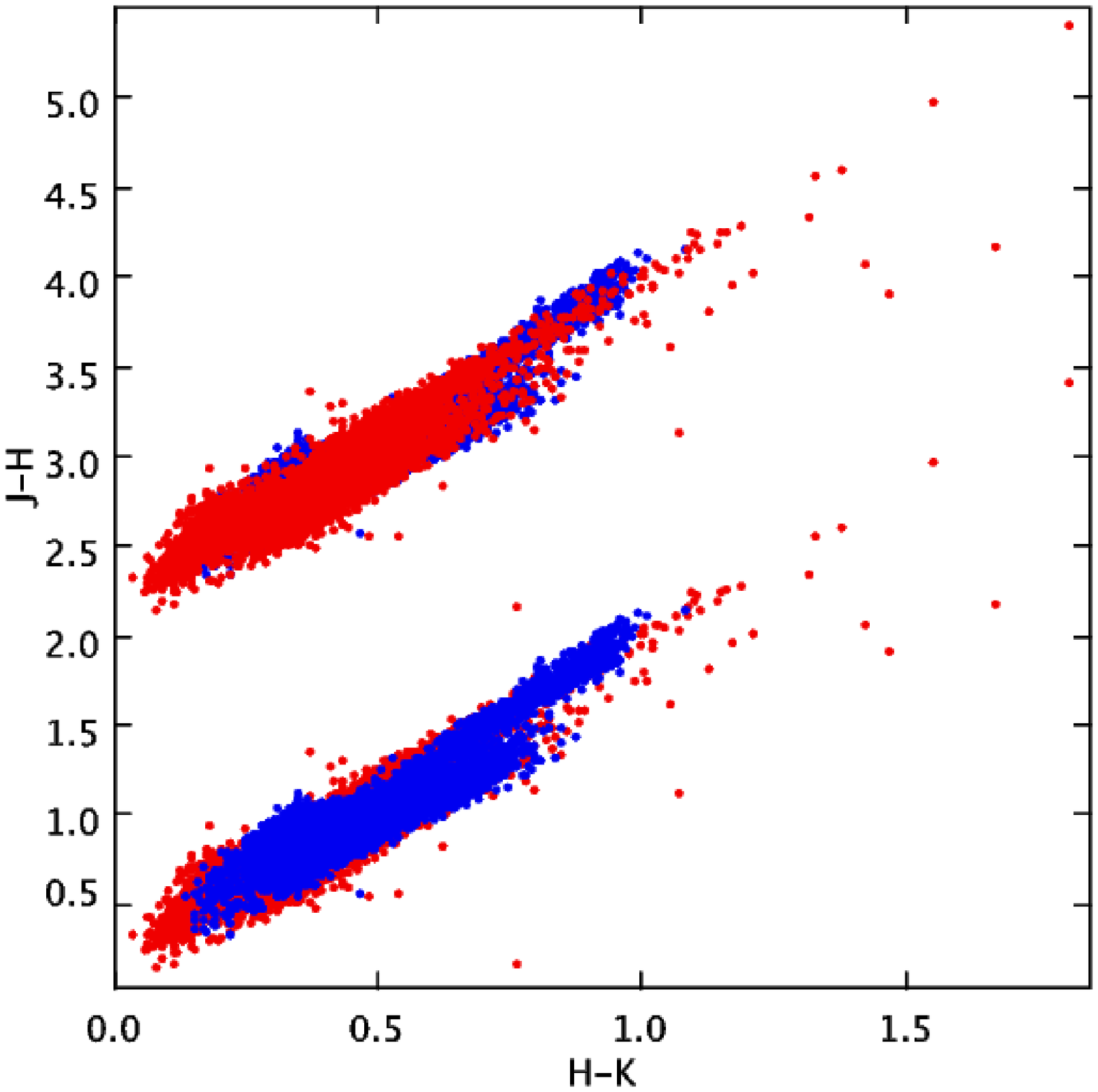}} \\
\resizebox{60mm}{!}{\includegraphics[angle=0]{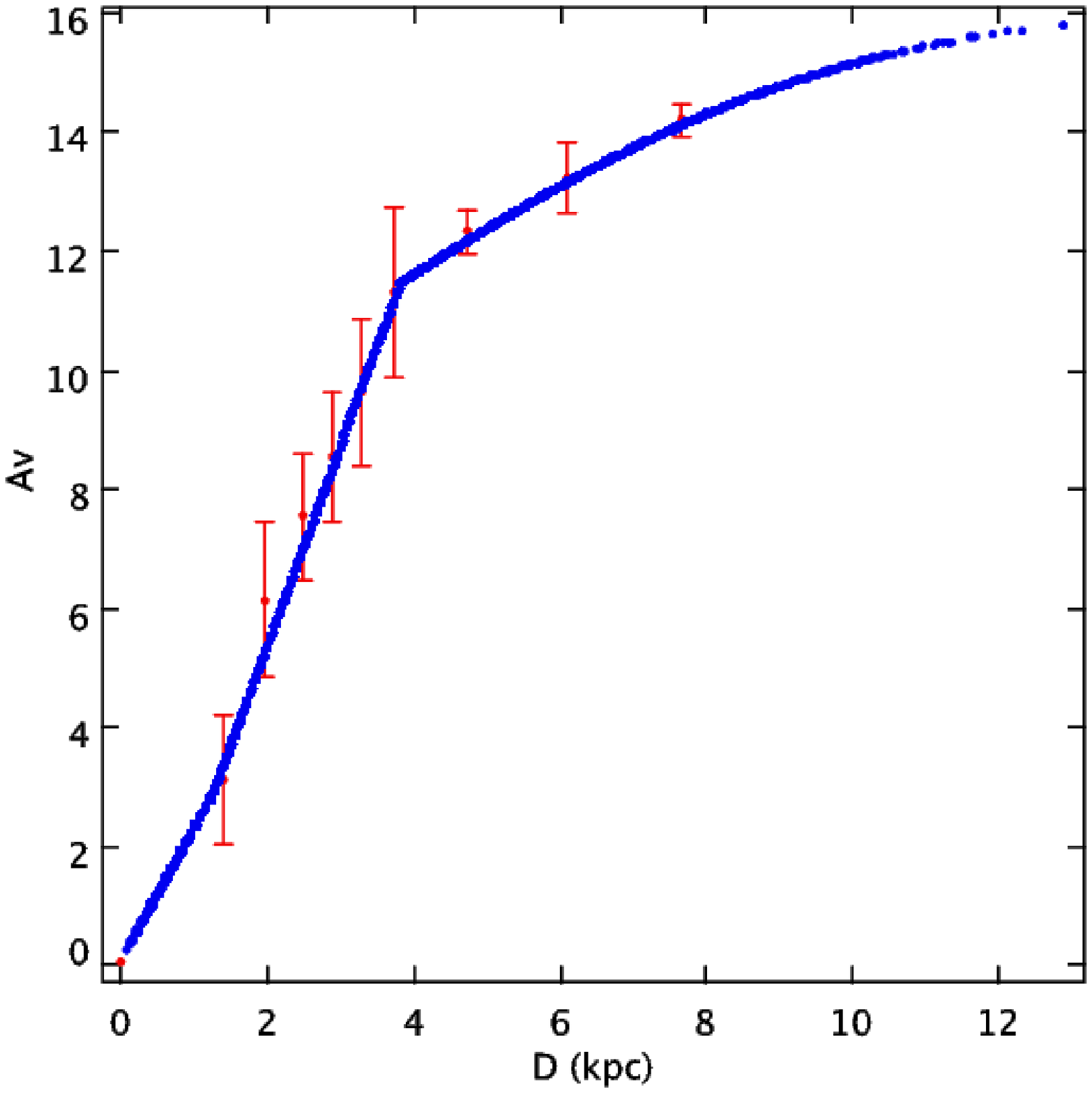}}&
      \resizebox{60mm}{!}{\includegraphics[angle=0]{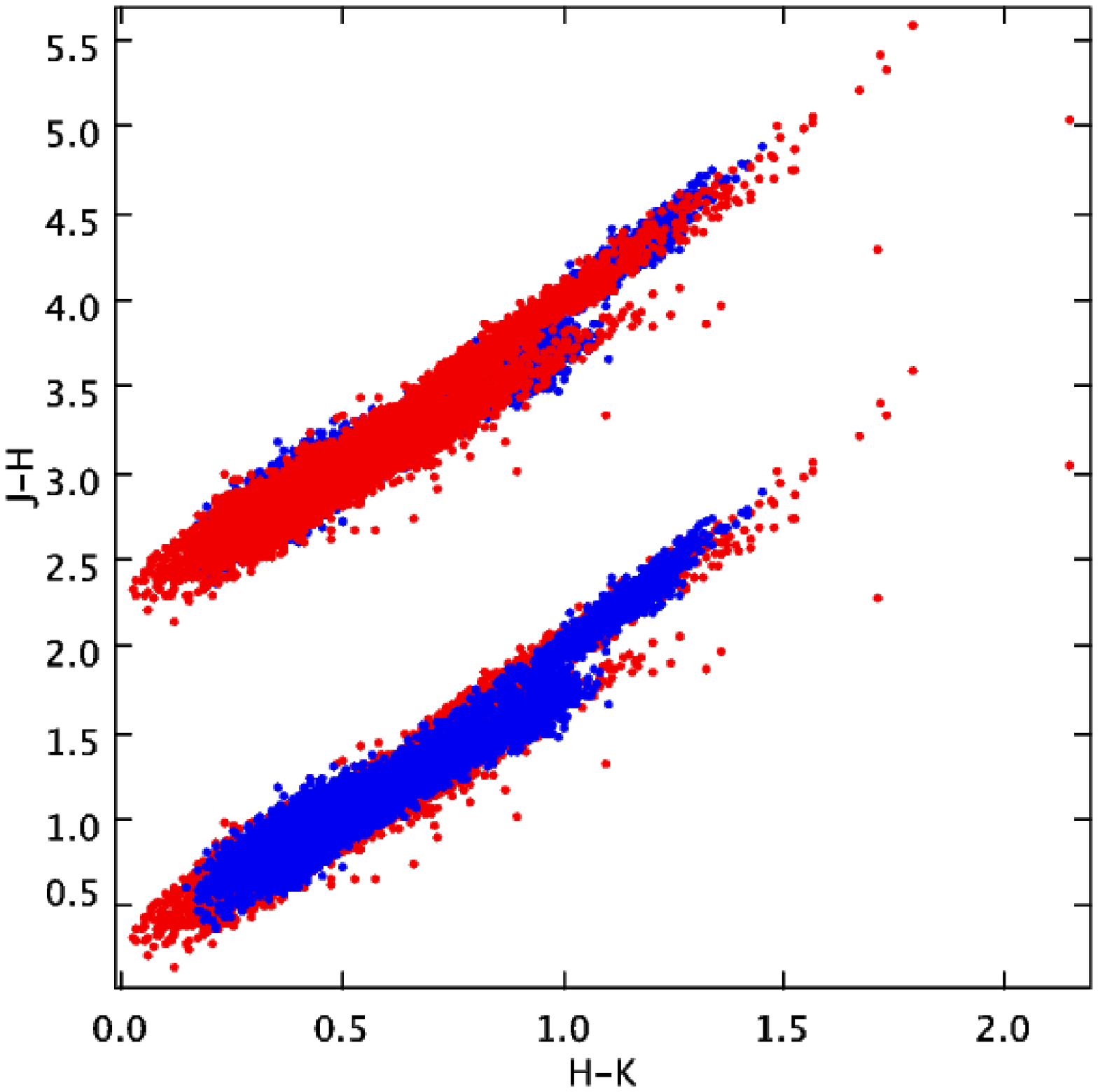}}\\
\resizebox{60mm}{!}{\includegraphics[angle=0]{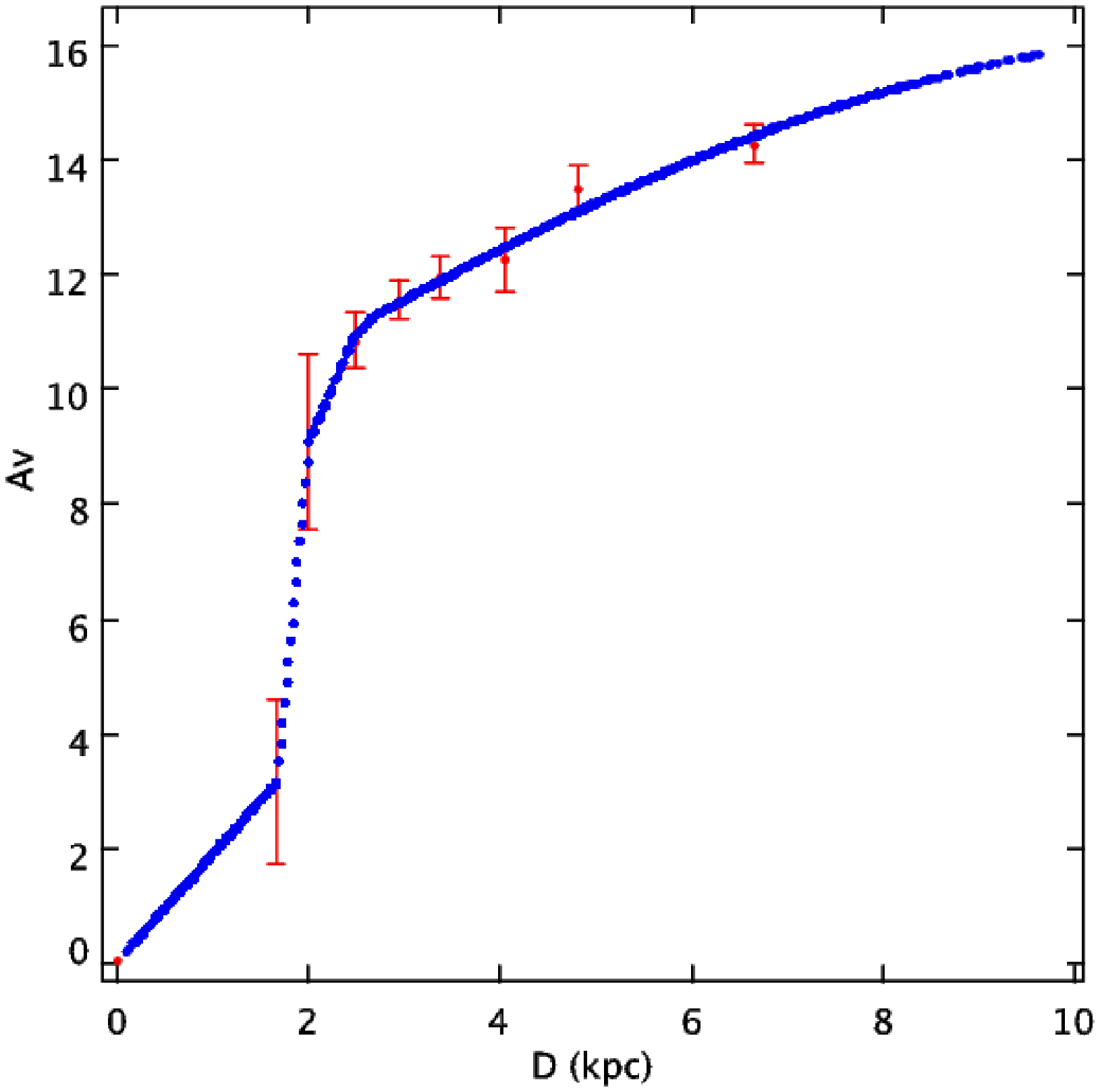}} &
      \resizebox{60mm}{!}{\includegraphics[angle=0]{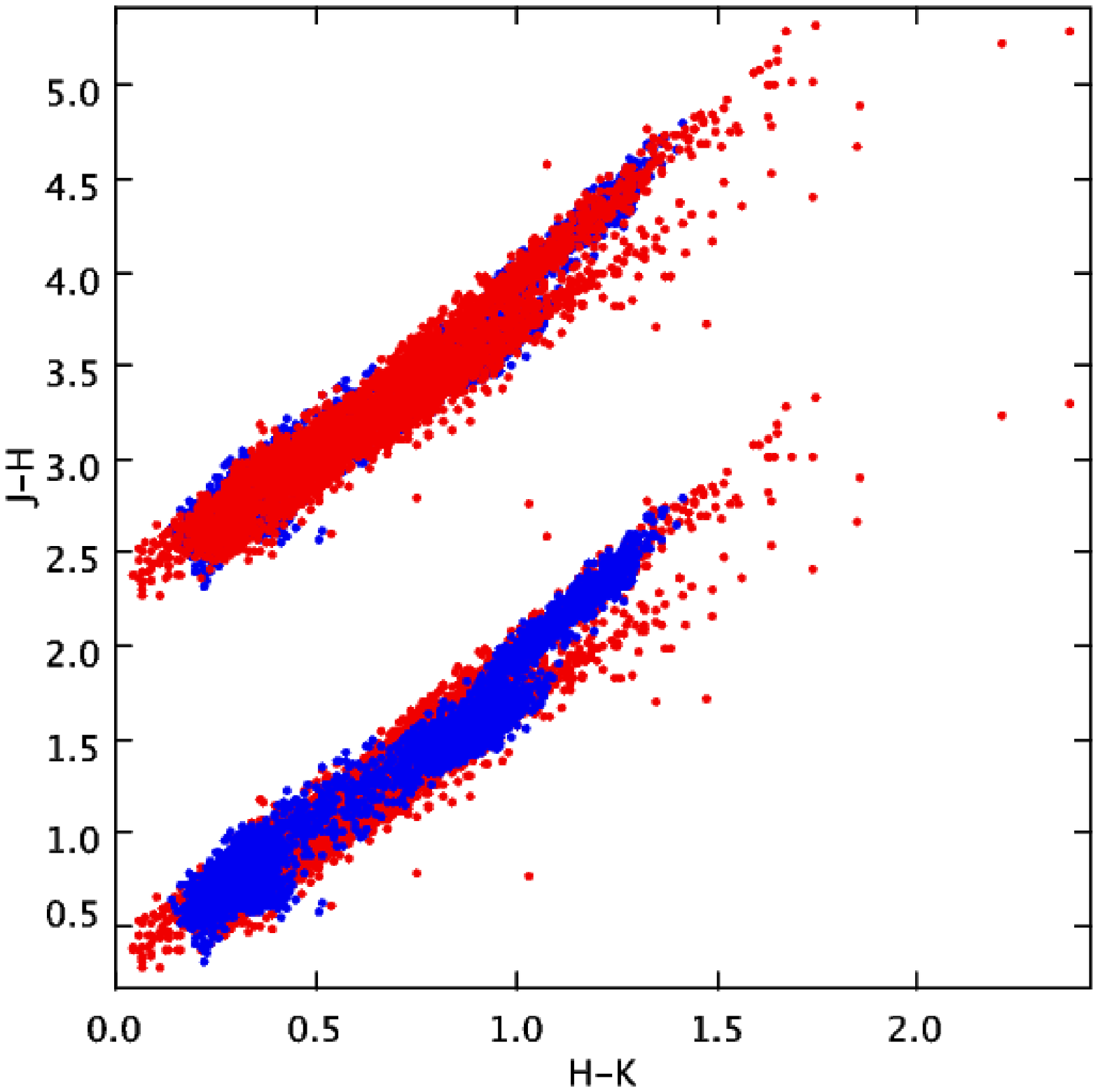}} 
    \end{tabular} 
\caption[]{\small a (top left): A distance-A$_V$ plot containing the extinction distribution model, lying within the constraints of the M06 data, that is used to derive each extinction power law for the region G69.5395. b (top right): A CCD created using UKIDSS data extracted from the region G69.5395 (red points) and $\Bes$ data (blue points) reddened with the derived law, $\alpha$=2.25. c (centre left): A distance-A$_V$ plot containing the extinction distribution model, lying within the constraints of the M06 data, that is used to derive each extinction power law for the region G77.9637. d (centre right): A CCD created using UKIDSS data extracted from the region G77.9637 (red points) and $\Bes$ data (blue points) reddened with the derived law, $\alpha$=2.20. e (bottom left): A distance-A$_V$ plot containing the extinction distribution model, lying within the constraints of the M06 data, that is used to derive each extinction power law for the region G82.1735. f (bottom right): A CCD created using UKIDSS data extracted from the region G82.1735 (red points) and $\Bes$ data (blue points) reddened with the derived law, $\alpha$=2.125.}
\label{fig:g69.5395}
\end{center}
\end{figure*}

\begin{figure*}
\begin{center}
    \begin{tabular}{cc}
      \resizebox{70mm}{!}{\includegraphics[angle=90]{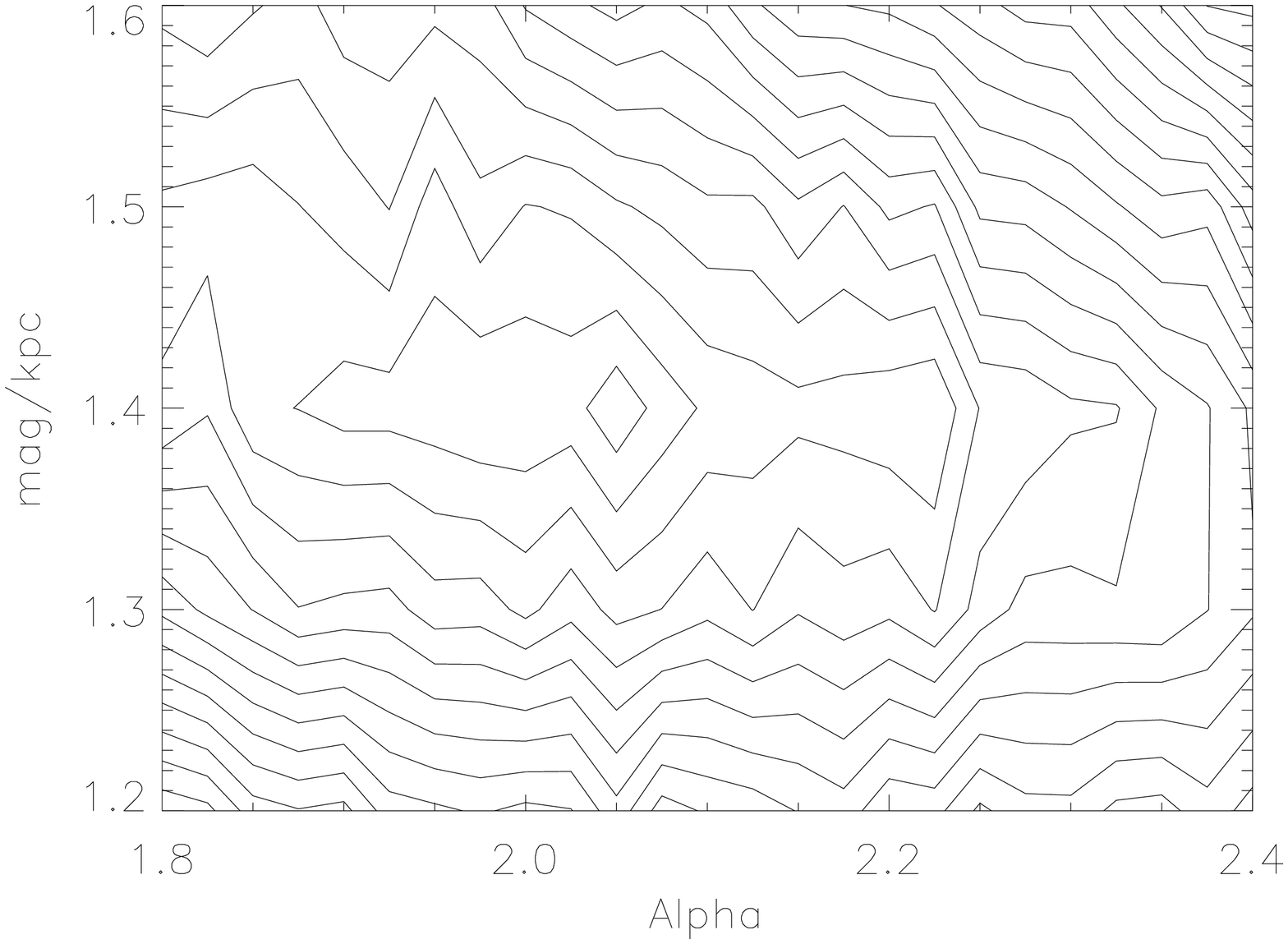}} &
\resizebox{55mm}{!}{\includegraphics[angle=0]{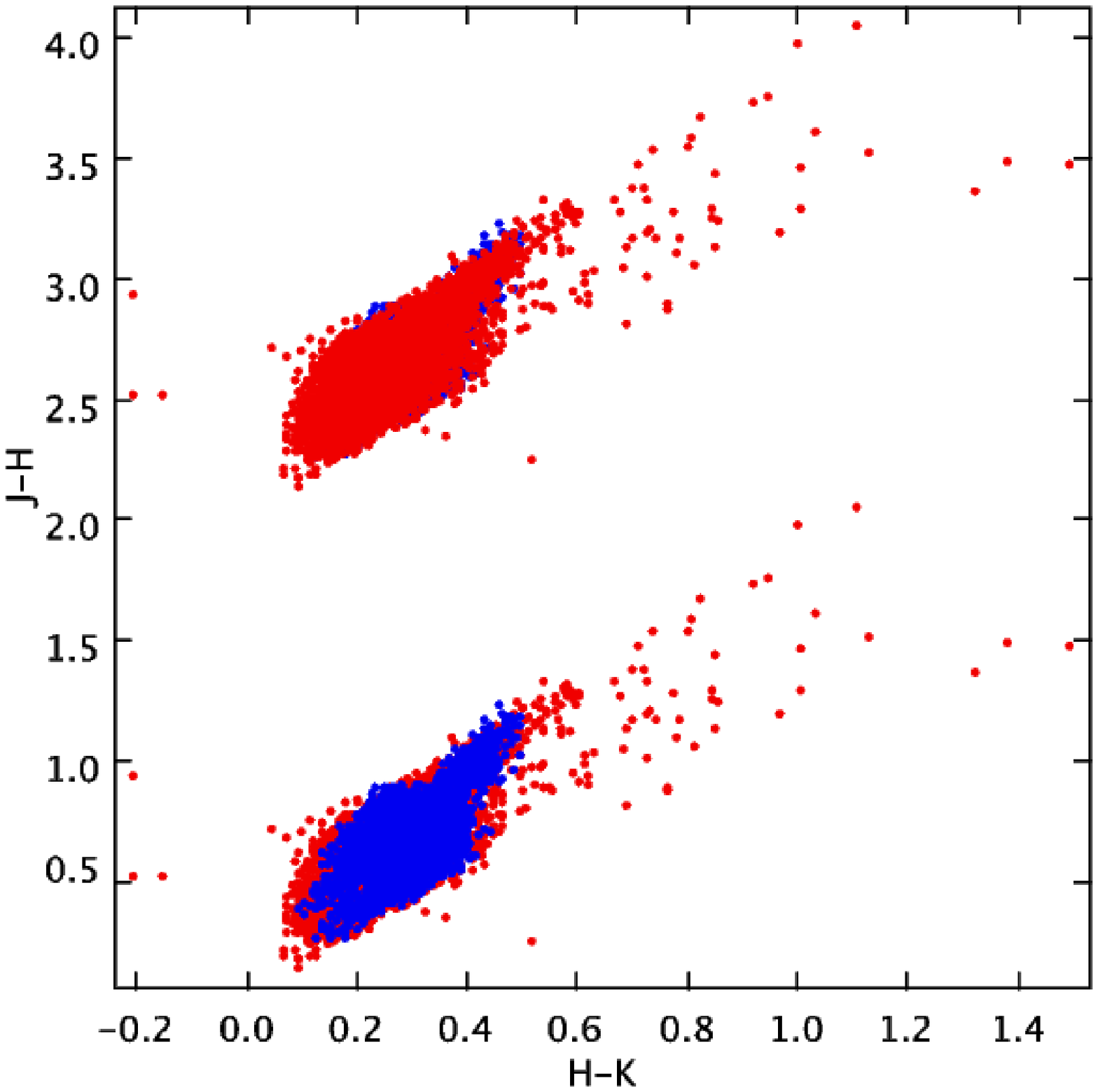}}
    \end{tabular} 
\caption[]{\small a (left): Simultaneous minimisation of L as the extinction model (mag/kpc) and value of $\alpha$ are varied. b (right): A CCD created using UKIDSS data extracted from the region G97.9978 (red points) and $\Bes$ data (blue points) reddened with the derived law, $\alpha$=2.05.}
\label{fig:g97.9978}
\end{center}
\end{figure*}

Results for the remaining 7 regions are presented in Table \ref{tabz:power}. Distance-A$_V$ plots including both the M06 data and any extinction distribution models used, along with the resultant CCDs created using UKIDSS data, are presented in Figures \ref{fig:g29.9564} to \ref{fig:g97.9978}. There is one exception  with the final region G97.9978. Due to the low amounts of extinction suffered and the low number of sources used to determine the fit, caused by the galactic position of the RMS source, it was not possible to constrain a value of $\alpha$ using 2MASS data and the extinction distribution model described by the M06 data. The M06 data appear to be an overestimate of the amount of extinction suffered, and due to the large area covered (90$\arcmin$x6$\arcmin$) it is probable that the M06 data was dominated by the high levels of extinction surrounding the RMS source. As the majority of data (excluding the highly reddened scattered sources) appear smooth and uniform we have performed, in the same manor as Section \ref{sec:extinctiondistribution}, a simultaneous minimisation of L as the smooth extinction model (mag/kpc) and value of $\alpha$ vary. Instead of a distance-A$_V$ plot, Fig.\ref{fig:g97.9978}a contains a contour plot similar to Fig.\ref{fig:varying_extinction}b. The contour lines are not as smooth as  Fig.\ref{fig:varying_extinction}b and this is reflected by the size of the derived errors.
\begin{table*}
\centering
TABLE \ref{tabz:power} \\
\small{DERIVED POWER LAWS FOR EACH REGION}\\
      \caption{ $^a$ see Section \ref{sec:modelcdd} for details. The presence of two kinematic distances highlights a kinematic distance ambiguity.}
\begin{tabular}{ c c c c c c c c c }
\hline
\hline
Region &Kinematic&Field&Extinction&Photometric&No Sources&Alpha&Error&Weighted\\
 &Distance&Size (lxb)&Model&System&&&&Mean \\
\hline
\hline
G48.9897-00.2992& 5.2 kpc	&60$\arcmin$x6$\arcmin$&	1.5m/k	&	UKIDSS	&6,638&		2.150	&	$^{+0.060}_{-0.085}$ \\
(HII region)&&&	&	2MASS	&	2,694	&	2.025	&	$^{+0.181}_{-0.081}$&2.088 $^{+0.095}_{-0.059}$	\\
&&&	1.2m/k	&	UKIDSS	&	 	&	2.200	&	$^{+0.059}_{-0.109}$	\\
&&&	&	2MASS	&	 	&	2.150	&	$^{+0.129}_{-0.204}$&2.175 $^{+0.071}_{-0.116}$	\\
&&&	1.8m/k	&	UKIDSS	&	 	&	2.125	&	$^{+0.085}_{-0.060}$	\\
&&&	&	2MASS	&	 	&	2.025	&	$^{+0.204}_{-0.129}$&2.075 $^{+0.111}_{-0.071}$	\\
&&&	Randomised	&	UKIDSS	&	 	&	2.225	&	$^{+0.106}_{-0.131}$	\\
&&&	Extinction$^a$	&		&		&		&	 \\
&&&	M06	&	UKIDSS	&	 	&	2.150	&	$^{+0.063}_{-0.063}$	\\
&&&		&	2MASS	&	 	&	2.050	&	$^{+0.180}_{-0.130}$&2.100 $^{+0.095}_{-0.072}$	\\
\hline
G29.9564-00.0174	&6.2/8.5 kpc&60$\arcmin$x6$\arcmin$&	M06	&	UKIDSS	&6,244	 	&	2.200	&	$^{+0.060}_{-0.085}$	\\
(HII region)&&&		&	2MASS	&3,870	 	&	2.200	&	$^{+0.154}_{-0.204}$&2.200 $^{+0.083}_{-0.111}$	\\
\hline
G42.1268-00.6224	&5.0/7.6 kpc&60$\arcmin$x6$\arcmin$&	M06	&	UKIDSS	&4,421	 	&	2.300	&	$^{+0.035}_{-0.110}$	\\
(HII region)&		&&&	2MASS	&3,024	 	&	2.075	&	$^{+0.179}_{-0.154}$&2.188 $^{+0.091}_{-0.095}$	\\
\hline
G53.6185+00.0376	&1.7/8.4 kpc&60$\arcmin$x6$\arcmin$&	M06	&	UKIDSS	&6,108	 	&	2.175	&	$^{+0.060}_{-0.085}$	\\
(YSO)&		&&&	2MASS	&3,237	 	&	2.050	&	$^{+0.215}_{-0.120}$&2.113 $^{+0.112}_{-0.074}$	\\
\hline
G69.5395-00.9754	&1.2/4.7 kpc&60$\arcmin$x6$\arcmin$&	M06	&	UKIDSS	&4,101	 	&	2.250	&	$^{+0.203}_{-0.228}$	\\
(HII region)&		&&&	2MASS	&2,774	 	&	2.050	&	$^{+0.277}_{-0.302}$&2.150 $^{+0.172}_{-0.189}$	\\
\hline
G77.9637-00.0075	&4.5 kpc&90$\arcmin$x6$\arcmin$&	M06	&	UKIDSS	&5,405	 	&	2.200	&	$^{+0.083}_{-0.108}$	\\
(HII region)&		&&&	2MASS	&2,211	 	&	2.150	&	$^{+0.204}_{-0.154}$&2.175 $^{+0.110}_{-0.094}$	\\
\hline
G82.1735+00.0792	&8.1 kpc&90$\arcmin$x6$\arcmin$&	M06	&	UKIDSS	&3,505	 	&	2.125	&	$^{+0.106}_{-0.131}$	\\
(YSO)&		&&&	2MASS	&1,489	 	&	2.275	&	$^{+0.178}_{-0.229}$&2.200 $^{+0.104}_{-0.132}$	\\
\hline
G97.9978+01.4688	&1.1 kpc&90$\arcmin$x6$\arcmin$&	1.4m/k	&	UKIDSS	&4,408	 	&	2.050	&	$^{+0.178}_{-0.328}$	\\
(YSO)&	&&&	2MASS	&2,650	 	&	1.925	&	$^{+0.178}_{-0.253}$&1.988 $^{+0.126}_{-0.207}$	\\
\hline
\hline
\end{tabular}
\label{tabz:power}
  \end{table*}

\section{Discussion}
\label{sec:discussion}
Creating an average of the derived extinction laws for all 8 regions results in a value of $\alpha$=2.14$^{+0.04}_{-0.05}$. All 8 regions agree to this average to within 1$\sigma$ uncertainties, suggesting that the value of $\alpha$ for the ISM is constant over the Galactic longitude range 27$^{\circ}$ $<$ l $<$ 100$^{\circ}$. Furthermore, there appears to be no change with differing Galactic latitude or total visual extinction. The slope towards each massive star forming region is steep so there is no sign of grain growth causing a flattening of the law. However the extinction is still likely to be dominated by the diffuse ISM in the large areas we analysed. \\
$\indent$The ability to test larger Galactic latitudes using this method will be limited, as at larger Galactic latitudes there is less total extinction and fewer background stars away from the Galactic Plane. As shown with the large error in the derived values of $\alpha$ for the regions G69.5395 and G97.9976, the method suffers from a limited amount of sources and extinction. \\
$\indent$\citet{mathis90} and \citet{draine03} present the results of several authors, many of whom have calculated values of $\alpha$ using colour excess ratio methods typically between 1.6-1.8. The main difference with this work is that the derived reddening law presented in this paper has been derived using filter wavelengths that evolve with the changing spectra of progressively reddened objects.\\
$\indent$\citet{moore05} explore the near-infrared extinction law in regions of high A$_V$ towards nine UCHII regions. They use hydrogen recombination lines to derive a value of $\alpha$ toward each region. Such a method does not use broad-band filters and is therefore not subject to biases caused by evolving effective wavelengths. The results of these nine UCHII regions give an average value of $\alpha$=1.59$\pm$0.30, although the range of values covered is large. Excluding results with $\tau$(Br$\gamma$) $>$ 3 (which corresponds to A$_V$$\sim$25, an approximate upper limit in extinction along the lines of sight presented in this paper), and only including distant sources ($>$3kpc), thereby making it likely that the extinction is dominated by the diffuse ISM, four sources remain. These data give an average value of $\alpha$=1.86$\pm$0.16. This result is consistent with the value derived in this paper at the 2$\sigma$ level. \\
$\indent$When colour excess ratios have been used to derive values of $\alpha$, the value depends on the specified wavelengths input into equation \ref{eq:no1}. It is unclear in the literature how exactly the filter wavelengths have been derived. Some authors have just referenced the photometric system and presented approximate wavelengths to specific filters \citep{whittet88,mart-whit90,mathis90}. When defining photometric systems, often it is either the isophotal or effective wavelengths, derived from specified spectra, that are given. Isophotal wavelengths are derived from the following equation \citep{golay74}: 
\begin{equation}
F({\lambda_{iso}}) = \frac{{\int}F({\lambda})S({\lambda}) d{\lambda}}{{\int}S({\lambda}) d{\lambda}},
  \label{eq:iso}
\end{equation}
where F($\lambda$) is the flux density of the specific spectra and S($\lambda$) is the throughput, taking into account the detector quantum efficiency, filter transmission, optical efficiency and the atmospheric transmission. The isophotal wavelength is then defined as the wavelength at which the monochromatic brightness (or magnitude) of the source spectrum equates to the weighted average flux across the filter band. As such it is therefore dependent upon the spectrum of the observed source.\\
$\indent$This problem is highlighted further when attempting to convert a given colour excess ratio to a value of $\alpha$. \citet{indebetouw05} derive a colour excess ratio E$_{H-K}$/E$_{J-K}$=0.35$\pm$0.02, towards the region referred to as l=284$^{\circ}$ (note, \citet{indebetouw05} only presented the derived colour excess ratios in their results and did not compute a value of $\alpha$). They have assumed that all of the sources used to calculate the colour excess ratio were K2III and so quote the isophotal wavelengths of the 2MASS filters, convolved with a K2III star, to be $\lambda _{J}$=1.240$\mu$m, $\lambda _{H}$=1.664$\mu$m and $\lambda _{K}$=2.164$\mu$m. Substituting these wavelengths and their colour excess ratio, that is equivalent to E$_{J-H}$/E$_{H-K}$=1.86$\pm$0.11, into equation \ref{eq:no1} yields a value of $\alpha$=1.81$\pm$0.20. As errors in the isophotal wavelengths were not provided they have not been included in the calculation. We have determined the effective wavelengths of the 2MASS filters using a CK04 K2III spectrum to be $\lambda _{J}$=1.244$\mu$m, $\lambda _{H}$=1.651$\mu$m and $\lambda _{K}$=2.159$\mu$m. Inputting these wavelengths into equation \ref{eq:no1} yields a value of $\alpha$=2.05$\pm$0.20. This value is now consistent with the average value derived in this paper. \\ 
$\indent$An attempt was made to derive an extinction law towards the region l=284$^{\circ}$ using the method described in this paper. However only 2MASS data were available as the Galactic location is outside the range of UKIRT. Furthermore, as the region covered 0.6$^{\circ}$ in Galactic latitude, thereby making it difficult to determine an extinction distribution model using M06 data, and the levels of total extinction were relatively low, it was not possible to constrain a value of $\alpha$ using 2MASS data. However, as \citet{indebetouw05} have assumed that the selected stars used to calculate the colour excess ratio are all of the same spectral type, the colour excess ratio can be considered to be an approximation to the curved reddening track (essentially a straight line fitted through the path of any reddening track with a gradient equal to the colour excess ratio). As \citet{indebetouw05} provided the range, in colour-colour space, over which the region was analysed, we have calculated the gradient of the straight line fitted to many CK04 K2III reddening tracks, created with varying values of $\alpha$, over the same range of data in colour-colour space. Therefore using a K2III reddening track, as opposed to a reddening vector to convert the colour excess ratio to a value of $\alpha$, we obtain $\alpha$=2.10$\pm$0.20. This is consistent with the value we derive. \\
$\indent$As a final test to highlight the problem of choosing the correct filter wavelengths, we have replaced the reddening tracks, used to derive an extinction law towards the region G48.9897-0.2992 using the smooth 1.5mag/kpc extinction distribution model, with a reddening vector. Results were derived for both photometric systems using, for 2MASS, the flat spectrum isophotal wavelengths from \citet{cohen03} and, for UKIDSS, the isophotal wavelengths from \citet{tok05}, and using effective wavelengths derived from a CK04 A0V spectrum. A comparison between the results is presented in Table 2. If previous authors have in the past used flat spectrum isophotal wavelengths to derive values of $\alpha$, then this could explain the reason why the value derived in this paper is much larger than those quoted in the literature.\\
\begin{table}
\centering
TABLE 2 \\ 
\small{Derived laws for G48.9897-0.2992 using a\\
1.5mag/kpc extinction distribution model}\\
\begin{tabular}{ c c c }
\hline
\hline
Photometric &Reddening&Alpha\\
system&method& \\
\hline
\hline
UKIDSS & Tracks & 2.150$^{+0.060}_{-0.085}$ \\
& $\lambda _{iso}$ & 1.950$^{+0.109}_{-0.059}$ \\
& $\lambda _{eff}$ & 2.150$^{+0.085}_{-0.060}$ \\
\hline
2MASS & Tracks & 2.025$^{+0.181}_{-0.081}$ \\
& $\lambda _{iso}$ & 1.750$^{+0.156}_{-0.106}$ \\
& $\lambda _{eff}$ & 1.975$^{+0.131}_{-0.131}$ \\
\hline
\hline
\end{tabular}
\label{tabz:eff_iso}
\end{table}
$\indent$It is of interest to compare our value of $\alpha$=2.14 to grain models. Data from \citet{draine84} show that the extinction suffered over NIR wavelengths (1.25$\mu$m to 2.20$\mu$m) is caused predominantly by the carbonaceous component rather than the silicate. \citet{draine84} use a power law distribution of grain sizes up to 0.25$\mu$m to fit the extinction law which resulted in a slope of 1.4 from 1.25$\mu$m to 2.20$\mu$m. \citet{kim94} use maximum entropy to derive the size distribution that showed an exponential cutoff at the upper end and gave a NIR slope of 1.8. If we examine the properties of single-sized graphite grains from \citet{draine1985} the NIR slope increases with increasing size from 1.55 for grains of radii 0.01$\mu$m, to 2.04 for 0.07$\mu$m and 2.39 for 0.10$\mu$m. Hence, a further weighting or extensions to larger carbon grains may be able to match the result derived in this paper. \cite{borghesi85} find an IR extinction law of $\sim$$\lambda^{-1}$ for amorphous carbon grains. This suggests that NIR extinction is not caused by amorphous carbon grains.
\section{Conclusion}
\label{sec:conclusion}
We have determined the slope of the NIR extinction power law, using two different photometric systems, along 8 different lines of sight, each with a massive star forming region towards the centre. Comparisons between colour-colour diagrams (CCDs) of UKIDSS and 2MASS data were made with CCDs of $\Bes$ Galactic model data. The $\Bes$ Galactic model, combined with the M06 extinction distribution models, reproduce the CCDs from UKIDSS and 2MASS data well. This enabled us to derive an extinction power law specific to each of the 8 different lines of sight. Each of the 8 extinction laws are consistent to within 1$\sigma$ uncertainties and show no variation with Galactic orientation or varying total visual extinction. An average extinction power law of $\alpha$=2.14$^{+0.04}_{-0.05}$, covers the Galactic longitude range 27$^{\circ}$ $<$ l $<$ 100$^{\circ}$. This value is considerably larger than previous values in the literature. It is necessary to consider the evolution of appropriate photometric effective wavelengths with changing observed spectra, particularly in deep fields when applying extinction correction techniques. In addition this means that straight reddening vectors commonly used in CCDs are in fact curved. The reddening tracks used in this paper can be downloaded at www.ast.leeds.ac.uk/RMS/ReddeningTracks/ . At the proof stage we were made aware of work similar to our own by \citet{Froebrich06}.

\section*{Acknowledgements}
We thank the anonymous referee for suggestions that significantly improved the paper. We acknowledge the invaluable assistance of Dr Jeremy Lloyd Evans for comments and suggestions towards the CCD fitting technique. This work is based in part on data obtained as part of the UKIRT Infrared Deep Sky Survey. This publication makes use of data products from the Two Micron All Sky Survey, which is a joint project of the University of Massachusetts and the Infrared Processing and Analysis Centre/California Institute of Technology, funded by the National Aeronautics and Space Administration and the National Science Foundation. We made use of the VizieR service (http://vizier.u-strasbg.fr/viz-bin/VizieR) to obtain 2MASS data and the M06 extinction distributions.

\bibliographystyle{mn2e}
\bibliography{Joey}{}

\end{document}